\renewcommand\i{\iota}

\newcommand{\diracslash}[1]{#1\llap{/\kern2pt}}

\newcommand{\be}{\begin{equation}}
	\newcommand{\ee}{\end{equation}}
\newcommand{\bea}{\begin{eqnarray}}
	\newcommand{\eea}{\end{eqnarray}}
\newcommand{\ba}[1]{\begin{array}{#1}}
	\newcommand{\ea}{\end{array}}

\newcommand{\bt}{\begin{tabular}}
	\newcommand{\et}{\end{tabular}}

\newcommand{\beas}{\begin{eqnarray*}}
	\newcommand{\eeas}{\end{eqnarray*}}

\RequirePackage{lineno}

 \documentclass[preprint,amsmath,amssymb,aps,a4]{revtex4-2}

\DeclareSymbolFont{rsfs}{U}{rsfs}{m}{n}
\DeclareSymbolFontAlphabet{\mathrsfs}{rsfs}

\usepackage{graphicx}
\usepackage[mathscr]{eucal}
\usepackage{multirow}
\usepackage{graphicx,epstopdf}
\usepackage{graphicx}
\usepackage{subcaption}
\usepackage{cancel}

\usepackage{setspace}
\usepackage[utf8]{inputenc}
\usepackage{csquotes}
\usepackage[english]{babel}
\usepackage{hyperref}
\usepackage{upgreek}
\hypersetup{
	colorlinks=true,
	linkcolor=blue,
	filecolor=magenta,     
	citecolor=blue, 
	urlcolor=blue,
}
\usepackage{cleveref}
\usepackage{caption}

\begin{document}
	\setstretch{1.5}
	\title{The in-medium mass and decay width of $\phi$ meson in dense and hot resonance matter} 
		\title{ $\phi$ meson properties in dense resonance matter at finite temperature} 
	\author{Manpreet Kaur}
	\email{ranapreeti803@gmail.com}
	\affiliation{Department of Physics, Dr. B R Ambedkar National Institute of Technology Jalandhar, 
		Jalandhar -- 144008, Punjab, India}

	\author{Arvind Kumar}
	\email{kumara@nitj.ac.in}
	\affiliation{Department of Physics, Dr. B R Ambedkar National Institute of Technology Jalandhar, 
		Jalandhar -- 144008, Punjab, India}

	\def\be{\begin{equation}}
		\def\ee{\end{equation}}
	\def\bearr{\begin{eqnarray}}
		\def\eearr{\end{eqnarray}}
	\def\zbf#1{{\bf {#1}}}
	\def\bfm#1{\mbox{\boldmath $#1$}}
	\def\hf{\frac{1}{2}}
	\def\kp{\zbf k+\frac{\zbf q}{2}}
	\def\km{-\zbf k+\frac{\zbf q}{2}}
	\def\hwo{\hat\omega_1}
	\def\hwt{\hat\omega_2}
	
\begin{abstract}
The effective mass and decay width of the $\phi$ meson in the isospin asymmetric hot and dense resonance matter are studied using the effective Lagrangian framework considering the $\phi K \bar K $ interactions at one-loop level. In addition to spin$-1/2$ octet baryons, we consider the effect of resonances $\Delta^{++,+,0,-}, \Sigma^{*\pm,0},\Xi^{*0,-}, \Omega^{-}$, on the properties of $\phi$ meson. The in-medium effects on the $\phi$ meson properties  are simulated through the effective masses of kaons and antikaons computed using the chiral SU(3) hadronic mean field model in the presence of resonance baryons. The loop integral appearing in the computation of $\phi$ meson self energies is regularized using the dipole form factor with a cutoff parameter. The presence of resonance baryons within the medium at finite temperature is observed to significantly modify the effective mass and decay width of $\phi$ mesons. Examining the $\phi$ meson masses and decay width within a dense medium is anticipated to be essential for understanding experimental results from heavy-ion collision experiments.		 
\end{abstract}
	\maketitle
    \newpage
\section{Introduction}
\label{intro}
The in-medium hadron modifications have been identified as a crucial component for interpreting the nonpertubative domain of quantum chromodynamics (QCD), which predicts several phases of matter at finite density and temperature \cite{Leupold:2009kz, hayano2010hadron}. As a result, essential investigations in this sector are both attractive and challenging. Due to the property of QCD confinement, at low temperature and moderate density, hadrons are treated as the degree of freedom in dense matter. However, at high temperature and density, these hadrons are anticipated to liberate quarks and gluons, forming the quark-gluon plasma (QGP) \cite{Maruyama:2007ss, Wang:2006fh}. 
    Gaining insight into the hadron-quark phase transition, where chiral symmetry is believed to be restored, requires examining the fundamental properties of hadrons and quarks as well as their interactions at finite densities. Understanding the thermodynamic properties of QCD plays a vital role in resolving natural events like compact stars and experimental findings from relativistic heavy-ion collisions (HICs)  \cite{hayano2010hadron, Leupold:2009kz, Tolos:2020aln, Pushkina:2004wa, Rafelski:2000by, Knospe:2017oin}. 
The heavy-ion collision experiments performed at very high center of mass energy such as Au-Au collisions at the relativistic heavy ion collider (RHIC), with $\sqrt{S_{NN}}=200$ GeV \cite{Harrison:2003sb,friman2011cbm,kumar2013star,odyniec2010rhic} and Pb-Pb collisions at large hadron collider (LHC) \cite{evans2007large, DiNezza:2024ctw}, with $\sqrt{S_{NN}}=2.76$ TeV, explore the QCD phase transition at high temperatures and low baryonic densities. Further insights are expected from forthcoming experiments, including the compressed baryonic matter (CBM) \cite{Agarwal:2023otg, senger2012compressed}, antiproton annihilation at Darmstadt (PANDA) \cite{Nilsson:2015aax, Herlert:2014uqa}, of the facility for antiproton and ion research (FAIR) at GSI \cite{rapp2010charmonium}, Japan proton accelerator research complex (J-PARC Japan) \cite{Sawada:2007gy, Hotchi:2017krx} and nuclotron-based ion collider facility (NICA) at Dubna, Russia \cite{Kekelidze:2017ghu, Kekelidze:2017tgp}, focusing on the physics of dense matter at finite baryonic densities.
\par

The hadrons produced in different HICs have a short lifetime and they decay with considerable probability within the strongly interacting environment. 
 In HICs, there are a number of reasons to be interested in measuring the in-medium modifications of light vector mesons, such as the large scale production of these mesons account for roughly 15\% of the total particle yields, their short lifetime, and significant electromagnetic decay properties. These properties make them valuable for examining the dense and hot hadronic matter through the dilepton pairs, which provide spectral knowledge of the vector mesons at the time of decay \cite{Milov:2008dd, Leupold:2009kz}.
    \par  
    QCD becomes highly challenging to solve in the non-perturbative low-energy regime because of the strong coupling of gluons and quarks. Consequently, researchers have explored alternative approaches to characterize hadrons within dense matter. The primary theoretical tools to explore the in-medium properties in the nonperturbative domain of QCD are (i) lattice QCD calculations and (ii) phenomenological models. 
    To investigate the QCD phase diagram in the sector of the high temperature and low baryonic density, one can use the lattice QCD simulation method, which is based on space-time grid points \cite{wilson1974confinement}. However, when the density of baryons rises, the sign problem with the fermion determinant causes lattice QCD to be inefficient, which leads to higher computational costs and lower productivity \cite{alexandru2022complex,bloch2009random,fukushima2007model,muroya2003lattice,takaishi2004hadronic}.
    Various phenomenological models and approaches have been developed to investigate hadronic matter outside the lattice QCD framework. These include quark meson coupling (QMC) model \cite{tsushima1999charmed, saito1994quark, Tsushima:1997cu},  Nambu–Jona–Lasinio (NJL)
 model \cite{Cao:2021rwx}, linear-sigma model (LSM) \cite{Bochkarev:1995gi}, hadron resonance gas (HRG) model \cite{Huovinen:2009yb}, Dyson-Schwinger equation approach \cite{Wadia:1980rb}, chiral perturbation theory (ChPT) \cite{pich1995chiral,scherer2003introduction,Waas:1996tw}, QCD sum rules \cite{hatsuda1992qcd,hayano2010hadron,kumar2011d}, chiral SU(3) hadronic and quark mean field model \cite{Zhang:1997ny,Wang:2004wja, Papazoglou:1998vr, Mishra:2009bp, Dexheimer:2008ax, kumar2020phi,chahal2024phi, Zschiesche:2003qq, Cruz-Camacho:2024odu,Kaur:2024cfm}, Polyakov quark meson (PQM) model \cite{Schaefer:2007pw,Herbst:2010rf,  Stiele:2013pma}, Polyakov loop extended NJL (PNJL) model and  the coupled channel approach \cite{ramos2000properties,Schaffner-Bielich:1999fyk,Lutz:1997wt,tolos2004properties,tolos2006d, tolos2008open, Wang:2024hwu,Liu:2023gla,Wilson:2023hzu}. Theoretical studies have broadened their reach by covering a variety of factors, including the exploration of non-extensive properties \cite{Rozynek:2009zh, Adel:2017uxi}, the impact of the external magnetic field \cite{Bali:2011qj,Bali:2012edd, Peterson:2023bmr}, the implications of finite system sizes \cite{Bhattacharyya:2015zka, Magdy:2015eda, Braun:2011iz}, and the examination of how electric fields influence the thermodynamics of strongly interacting systems \cite{Ruggieri:2016lrn, Ruggieri:2016xww}.
     \par
 In HICs, even at relatively low energies, several kinds of mesons, non-strange and strange baryons, in addition to pions, nucleons, and resonances, are produced.
The production of strangeness in these collisions 
and neutron stars is of considerable interest because they provide a way to answer many fundamental questions of nuclear and hadronic physics connected to the many concepts of QCD \cite{Senger:2010zz, Yong:2023aal, Tolos:2020aln}.  Due to the conservation of strangeness in hadronic reactions, the production of strange hadrons is essential. The production of open strangeness occurs through the creation of kaon-hyperon combinations or the formation of kaon and antikaon pairs $K \bar K$. It is possible to produce the hidden strangeness through $\phi$ ($s \bar s$) meson. 
There are two ways to measure the $\phi$ meson. The first involves analyzing the dilepton decay ($
\phi \to e^+ e^-$ or $
\phi \to \mu^+ \mu^-$), which offers undistorted information about $\phi$ meson's behavior within the medium because of negligible interaction of the dileptons with baryons and mesons. However, this approach is challenging because of the low branching ratio of $\phi$ meson to dileptons. The second method examines $K \bar K$ pairs resulting from $\phi$ meson decay with a high branching ratio \cite{Song:2022jcj}. The $\phi$ meson dominantly decays into the kaon-antikaon pair due to the suppression of $\phi$ decay via the $\pi$ meson, a consequence of the Okubo-Zweig-Iizuka (OZI) rule \cite{Okubo:1963fa,iizuka1966systematics,lipkin1984theoretical}.



Numerous theoretical \cite{asakawa1994phi,hatsuda1992qcd,hatsuda1996light,blaizot1991varphi, kuwabara1995varphi,oset2001test,bhattacharyya1997medium,klingl1997current,Cobos-Martinez:2017vtr,Cobos-Martinez:2017woo} and experimental \cite{Leupold:2009kz,hayano2010hadron,muto2007evidence,mibe2007measurement, qian2009extraction,ishikawa2005phi} investigations have been conducted to examine the properties of the $\phi$ meson within strongly interacting hadronic matter. In HICs, the $\phi$ meson has been experimentally observed in both $e^+ e^-$ and $K^+ K^-$ invariant mass spectra, ranging from GSI (1-2 GeV) to LHC (2-5 TeV) energies \cite{Sako:2024oxb}. 
However, the mass shift of $\phi$ meson was found to be relatively negligible compared to their lifetime, i.e., a large considerable broadening in decay width of $\phi$ has been noted. Such broadening effect results because of the change in the hadronic decay channels and collision processes due to
hadron-hadron scattering interactions \cite{Leupold:2009kz, hayano2010hadron}.  
The E325 experiment at KEK \cite{muto2007evidence} was an earlier experiment that investigated the in-medium modification of $\phi$ meson by measuring its decay into $e^+ e^-$ and $K \bar K$ pairs in $p + A$ collisions, 
observing  no modification of invariant mass spectra of $K \bar K$ at nuclear saturation density. In Ref. \cite{ishikawa2005phi},  SPring8 collaboration, which employed a Glauber-type multiple scattering theory and the A-dependence of $\phi$ photo-production yields, observed a significant $\phi N$ cross section in the medium, leading to a decay width of 35 MeV, which aligns with the experimental findings. 
Further, to improve the statistics significantly, the J-PARC E16 and E88 experiments \cite{Aoki:2024ood,Sako:2024oxb} have been proposed as the successors to the E325 experiment, to explore the origin of QCD mass by examining the spectral change of $\phi$ meson within the nuclear medium \cite{Sako:2024oxb}. 


\par
From a theoretical point of view, various approaches such as QCD sum rules \cite{asakawa1994phi,hatsuda1992qcd,hatsuda1996light}, the Nambu-Jona-Lasinio model \cite{blaizot1991varphi}, effective Lagrangian approaches  \cite{kuwabara1995varphi,oset2001test,bhattacharyya1997medium,klingl1997current} have been established, which claim a decrease of $\phi$ meson mass in the dense nuclear medium. 
In addition to this, the coupled-channel approach examines kaon-hyperon interaction, incorporating multiple channels and resonances \cite{chiang2004dynamical,ramos2000properties,lutz1998nuclear}.
 These results are further used to determine the modification of in-medium properties of $\phi$ meson \cite{Klingl:1997tm}. In Ref. \cite{Gubler:2014pta}, Gubler and Ohtani report the in-medium results of $\phi$ meson using the QCD sum rule approach. Their findings revealed that the mass shift of $\phi$ within nuclear matter is closely linked to the nucleon strangeness content $\sigma_{SN}$ and based on the value of $\sigma_{SN}$, the $\phi$ meson may experience either an increase or decrease in the mass at nuclear saturation density. Various relativistic transport models have also been employed to examine $\phi$ meson production in HICs \cite{Song:2022jcj, Chung:1997mp, Pal:2002aw}. For instance, the Parton-Hadron-String Dynamics (PHSD) transport model focuses on interactions between baryon-baryon and meson-baryon. The PHSD calculations, which incorporate medium-modified effects like the collisional broadening of the spectral function of the $\phi$ meson, successfully replicated the experimentally observed $K/ \phi$ ratio \cite{Song:2022jcj}.
As discussed in Ref. \cite{Chung:1997mp, Pal:2002aw}, the reduction in $\phi$ meson yield is primarily due to a rise in its decay width resulting from the reduction in kaon-antikaon masses. 
The study of in-medium masses of $\phi$ mesons is also important to explore the possible formation of their bound states with different nuclei \cite{Cobos-Martinez:2017woo,Paryev:2022zkt,Kumar:2024lfa,Mondal:2025qxm}.
Moreover, recent studies in Refs. \cite{Tu:2021sxx, Wu:2023ecx} have examined how  $\phi$ meson affect neutron stars and hyperon stars within the framework of the relativistic mean field model.
\par
Several studies have addressed resonance production in different collision systems measured by ALICE at LHC \cite{Knospe:2017oin, Song:2020bnc, Lomker:2024ctm, Rosano:2022wlj, Kiselev:2022tyj}. These studies give information about the final state of hadronic reactions.
The chiral perturbation theory \cite{Bedaque:1995pa}, NJL and PNJL model \cite{Torres-Rincon:2015rma}, and thermal QCD sum rules \cite{Xu:2015jxa, Azizi:2016ddw} have been used to examine temperature effects on decuplet baryon masses. QCD sum rules were employed to analyze the in-medium masses of decuplet baryons in symmetric nuclear matter \cite{Azizi:2016hbr}. In Ref. \cite{Ouellette:1997ip}
self-energy of decuplet baryons within nuclear matter has been studied using the chiral perturbation theory. The chiral SU(3) quark mean field model was utilized to study in-medium masses and magnetic moments of decuplet baryons in nuclear and hyperonic medium \cite{Kumar:2023owb, Ryu:2008st, Singh:2017mxj, Singh:2020nwp}. Dense resonance matter properties at zero temperature were investigated using the chiral SU(3) model \cite{Zschiesche:2000ew}. 
Heavy baryon resonance properties in nuclear matter have also been studied \cite{Lenske:2018bvq, Singh:2017mxj}. The influence of resonance baryons, particularly $\Delta$ resonances, on compact star mass-radius relations has been explored using various approaches \cite{Glendenning:1982nc, Oliveira:2019xni, Raduta:2020fdn, Dexheimer:2021sxs}.
\par
In this work, we employ the chiral SU(3) hadronic model to investigate in-medium $\phi$ meson mass and decay width in isospin asymmetric resonance matter. 
In the $\phi$ meson decay, we examine the $K\bar{K}$ loop contributions by solving the effective Lagrangian of the $\phi{K}\bar{K}$ interaction. Furthermore, with the help of regularization techniques, we evaluate loop integral for in-medium self-energy.
This paper is organized in the  following manner: 
 Secs. \ref{SEC_IIA} and \ref{SEC_IIB}  address   the chiral SU(3) hadronic mean field model \cite{Papazoglou:1998vr} used to obtain the in-medium scalar and vector fields in dense resonance matter and subsequently the in-medium masses of kaons and antikaons.
  The theoretical perspective to estimate the in-medium modifications of $\phi$ meson is explored in Sec. \ref{SEC_IIC}. Sec. \ref{results} provides the results and discussions and we conclude with the summary in Sec. \ref{summary}.

\section{Methodology}
\label{math}

The chiral SU(3) hadronic model is applied in the current study to probe the influence of isospin asymmetry and strangeness fraction on the thermodynamic properties of hadronic matter by mesonic exchange ($\sigma, \zeta$, $\delta$, $\omega, \rho$ and $\phi$). Furthermore, we use a self-consistent Lagrangian technique to compute the in-medium properties of the $\phi$ meson. In the following subsections, we concisely explain the basic formalism used to derive the results for $\phi$ meson properties.
\subsection{The chiral SU(3)$_L$ $\times$ SU(3)$_R$ hadronic model}
\label{SEC_IIA}
The chiral SU(3)$_L$ $\times$ SU(3)$_R$ model has been successfully employed to analyze hadronic matter, finite-sized nuclei, hypernuclei, and neutron stars  in which the baryonic chemical potential becomes large enough to produce particles beyond the lowest SU(2) multiplet of nucleons, including hyperons ($\Lambda, \Sigma^{\pm,0}, 
 \Xi^{-,0}$) and other resonance states ($\Delta^{++,+,0,-}, \Sigma^{*\pm,0}, \Xi^{*0,-}, \Omega^{-}$). This model employs a flavor-SU(3) framework, incorporating octet and decuplet baryons along with strange mesons as fundamental elements within the primary hadronic SU(3) multiplets.
The various terms of effective hadronic Lagrangian density of this model are given as
\begin{equation}
\label{gen_L}
\mathcal{L} 
  = \mathcal{L}_{\text{kin}} + \mathcal{L}_{{int}} +
	\mathcal{L}_{scal}+ \mathcal{L}_{\text{vec}} + \mathcal{L}_{\text{SB}}.  
\end{equation}
This equation arises from the relativistic quantum field theory of baryons and mesons, which incorporates essential QCD features, including broken scale invariance \cite{Papazoglou:1998vr, weinberg1968nonlinear} and the non-linear realization of chiral symmetry \cite{bardeen1969some}. 
The term $\mathcal{L}_{kin}$ in Eq. (\ref{gen_L}) is the kinetic energy term for different particles, and the term $\mathcal{L}_{int}$ includes the interaction of baryons with various mesons.  As discussed before, the interaction of octet and decuplet baryons is considered through the exchange of   scalar fields, $\sigma, \zeta$ and $\delta$ and vector fields, $\omega, \rho$ and $\phi$. Medium-range attractive interactions between hadrons are influenced by scalar fields, while vector fields affect short-range repulsive interactions.
The interactions occurring between baryons via scalar mesonic fields are responsible for generating the baryon masses. This occurs through the coupling of baryons to two distinct scalar quark condensates, (i) the non-strange $\sigma$, which is composed of up and down quarks ($ \sim \langle \bar{u} u + \bar{d} d \rangle$)
 and (ii) the strange $\zeta$, which is composed of strange quarks $ \sim \langle \bar s s \rangle$. Additionally, to address the isospin asymmetry within the medium and the QCD trace anomaly properties, this model incorporates the scalar isovector field $\delta$ and the dilaton field $\chi$, respectively. 
 In Eq.~(\ref{gen_L}), the third term $\mathcal{L}_{scal}$ and the fourth term $\mathcal{L}_{vec}$  of the Lagrangian density represent the self-interaction for scalar and vector mesons, respectively.  The last term, $\mathcal{L}_{SB}$, corresponds to the explicit breaking of chiral symmetry, which serves to eliminate the Goldstone bosons. Detailed explanations of interaction Lagrangian density terms can be found in the literature \cite{Papazoglou:1998vr,Cruz-Camacho:2024odu, Kaur:2024cfm}. 
 \par
 In order to examine the properties of hadronic matter in a nonperturbative domain, we utilize the mean-field approximation. This method involves substituting the quantum field operator with their anticipated classical expectation values, i.e., $\Phi(x) = \langle \Phi \rangle + \delta \Phi \rightarrow \langle \Phi \rangle \equiv \Phi $ for scalar fields and $	V_\mu(x) = \langle V \rangle \delta_{0\mu} + \delta V_\mu \rightarrow \langle V_0 \rangle \equiv V $ for vector fields as these fields are expected to exhibit minimal fluctuations in high-density environments
 \cite{Papazoglou:1998vr}. 
  The thermodynamic potential per unit volume $\frac{\Omega}{V}$ for isospin asymmetric resonance medium is calculated under this approximation and is expressed by the following equation
\begin{align}
    \frac{\Omega}{V}&=-\frac{\gamma_i T}{(2 \pi)^3} \sum_i \int d^3 k\left\{\ln \left(1+e^{-\beta\left[E_i^*(k)-\mu_i^*\right]}\right)+\ln \left(1+e^{-\beta\left[E_i^*(k)+\mu_i^*\right]}\right)\right\}\nonumber\\
&-\mathcal{L}_{M} -V_{vac} .
\label{Eq_TP}
\end{align}

Here, the summation is over the baryons of the medium, i.e., $i=p,n,\Lambda,\Sigma^{\pm,0}, 
\Xi^{-,0},\Delta^{++,+,0,-},\nonumber \\ \Sigma^{*\pm,0},\Xi^{*0,-}, \Omega^{-}$.  $\gamma_i$ denotes the spin degeneracy factor, which equals $2$ for octet baryons and $4$ for decuplet baryons, and $\mathcal{L}_{M} = \mathcal{L}_{v e c}+\mathcal{L}_0+\mathcal{L}_{S B} $.
Also, \(E_i^*(k) = \sqrt{k^2 + m_i^{*2}}\)
and effective baryon mass, $m_i^*$, is calculated using relation
\begin{equation}
 m_i^*=-\left(g_{\sigma i} \sigma+g_{\zeta i} \zeta+g_{\delta i} \tau_{3i} \delta\right)+ m_{i0},
 \label{mass_B}
\end{equation}
 where  $g_{\sigma i}$, $g_{\zeta i}$, and $g_{\delta i}$ are couplings of $\sigma$, $\zeta$, and $\delta$ fields, respectively, with different baryons and  the term $m_{i0}$ is fitted to obtain  the baryon vacuum masses.
 Also, $\tau_{3i}$ has values $+1$ for $p, \Sigma^{+}, \Xi^{0}, \Delta^{+},\Sigma^{*+}, \Xi^{*0}$, $-1$ for $n, \Sigma^-, \Xi^{-}, \Delta^{0}, \Sigma^{*-}, \Xi^{*-}$, $0$ for $\Lambda, \Sigma^{0},\Sigma^{*0} $, $+3$ for $\Delta^{++}$ and $-3$ for $\Delta^{-}$ baryons.
The effective chemical potential of baryons is defined as \(\mu_i^* = \mu_i - g_{\omega i} \omega - g_{\phi i} \phi - g_{\rho i} \tau_{3i} \rho\), where \(g_{\omega i}\), \(g_{\phi i}\), and \(g_{\rho i}\) serve as the coupling constants between the baryons and \(\omega\), \(\phi\), and \(\rho\) fields, respectively. 
 These vector coupling constants are adjusted to match nuclear saturation properties, such as the binding energy ($B/A=-16$ MeV) at $\rho_0=0.15$ fm$^{-3}$ for symmetric nuclear matter.
The term $V_{vac}$ in Eq.~(\ref{Eq_TP}) is subtracted to ensure that the vacuum energy is zero.
The Eq. (\ref{Eq_TP}) is subjected to minimization with respect to the mesonic fields through the following relations

\begin{eqnarray}
	\frac{\partial \Omega}{\partial \sigma} = 
	\frac{\partial \Omega}{\partial \zeta} =
	\frac{\partial \Omega}{\partial \delta} =
	\frac{\partial \Omega}{\partial \chi} =
	\frac{\partial \Omega}{\partial \omega} =
	\frac{\partial \Omega}{\partial \rho}  =
    \frac{\partial \Omega}{\partial \phi}  =
	0.
	\label{eq:therm_min1}
\end{eqnarray} 
 At finite values of baryon density, $\rho_{B}$, temperature, $T$, isospin asymmetry, $I_a$, and strangeness fraction, $f_{s}$ of the medium, the non-linear equations obtained through above relations for both the scalar and vector fields are solved. 
 Additionally, the vector and scalar densities of the baryons, denoted by the symbols $\rho_i^v$ and $\rho_i^s$, respectively, are expressed as 
 \begin{eqnarray}
\rho_{i}^{v} = \gamma_{i}\int\frac{d^{3}k}{(2\pi)^{3}}  
\Bigg(\frac{1}{1+\exp\left[\beta(E^{\ast}_i(k) 
-\mu^{*}_{i}) \right]}-\frac{1}{1+\exp\left[\beta(E^{\ast}_i(k)
+\mu^{*}_{i}) \right]}\Bigg) ,
\label{rhov0}
\end{eqnarray}
and
\begin{eqnarray}
\rho_{i}^{s} = \gamma_{i}\int\frac{d^{3}k}{(2\pi)^{3}} 
\frac{m_{i}^{*}}{E^{\ast}_i(k)} \Bigg(\frac{1}{1+\exp\left[\beta(E^{\ast}_i(k) 
-\mu^{*}_{i}) \right]}+\frac{1}{1+\exp\left[\beta(E^{\ast}_i(k)
+\mu^{*}_{i}) \right]}\Bigg) .
\label{rhos0}
\end{eqnarray}
For the variables $I_a$ and $f_{s}$, their definitions are $-\frac{\sum_i I_{3i} \rho^{v}_{i}}{\rho_{B}}$ and $\frac{\sum_i \vert S_{i} \vert \rho^{v}_{i}}{\rho_{B}}$, respectively, with $\rho_B$ as the total baryonic density of the  medium under consideration.



\subsection{In-medium masses of $K$ and $\bar K$ mesons}
\label{SEC_IIB}

To evaluate $\phi$ meson properties, it is essential first to calculate in-medium kaon (antikaon) masses, which are influenced by how  $K$ ($\bar K$) interact with the surrounding resonance environment. In this section, we calculate medium-modified masses of kaons, $K$ $\left(K^+, K^0\right)$, and antikaons, $\bar K \left(K^-, \bar{K^0} \right)$, in the isospin asymmetric dense resonance medium via the dispersion relation.    
The Lagrangian density that reflects the interaction of $K$ ($\bar K$) with octet and decuplet baryons can be written as \cite{Kaur:2024cfm,Mishra:2008dj}
\begin{equation}
\label{Eq_kBD_gen1}
\mathcal L _{KB} =\mathcal{L}_{BW}+ \mathcal L _{mass} + {\mathcal L}_{{\mathrm{1st\, range\, term}}} + {\mathcal L }_{d_1}^{BM} + {\mathcal L }_{d_2}^{BM} .
\end{equation}

The first term, $\mathcal{L}_{BW}$, is described as a vectorial interaction term, which is derived from the kinetic term of interaction Lagrangian of Eq. (\ref{gen_L}) and can be expressed as $\mathcal{L}_{\text{BW}} = i \operatorname{Tr} \left( \bar{B} \gamma^{\mu} D_{\mu} B \right)-i \bar{T}^\mu \mathcal{\cancel{D}} {T}_\mu$, where $B$ and $T^\mu$ are octet and decuplet baryon fields, respectively and $D_{\mu}$ denotes covariant derivative. The second term is a mass term, which originates from explicit symmetry breaking, while the third term is a result of the kinetic terms of pseudoscalar mesons in the chiral effective Lagrangian. The ${\mathcal L }_{d_1}^{BM}$ and ${\mathcal L }_{d_2}^{BM} $ terms are referred to as range terms, obtained from the chiral model's baryon-meson interaction Lagrangian
\cite{Mishra:2008dj,Mishra:2008kg}.
Equation (\ref{Eq_kBD_gen1}) can be expanded to the following
\cite{Kaur:2024cfm,Mishra:2008dj}
\begin{eqnarray}
\label{Eq_kBD_gen2}
\mathcal L _{KB} & = & -\frac {i}{4 f_K^2} \Big [\Big ( 2 \bar p \gamma^\mu p
+\bar n \gamma ^\mu n -\bar {\Sigma^-}\gamma ^\mu \Sigma ^-
+\bar {\Sigma^+}\gamma ^\mu \Sigma ^+
- 2\bar {\Xi^-}\gamma ^\mu \Xi ^-
- \bar {\Xi^0}\gamma ^\mu \Xi^0 \nonumber \\
&+& 3( 3 \bar {\Delta^{++}} \gamma^\mu \Delta^{++} +2 \bar {\Delta^{+}} \gamma^\mu \Delta^{+} + \bar {\Delta^{0}} \gamma^\mu \Delta^{0}- \bar {\Sigma^{*-}}\gamma ^\mu \Sigma ^{*-}+ \bar {\Sigma^{*+}}\gamma
^\mu \Sigma ^{*+}
\nonumber\\
& -&2 \bar {\Xi^{*-}}\gamma ^\mu \Xi ^{*-} - \bar {\Xi^{*0}}\gamma ^\mu \Xi ^{*0} - 3\bar {\Omega^-}\gamma ^\mu \Omega ^-)\Big)
\nonumber \\
& \times & \Big(K^- (\partial_\mu K^+) - (\partial_\mu {K^-})  K^+ \Big ) \nonumber \\
& + & \Big ( \bar p \gamma^\mu p
+ 2\bar n \gamma ^\mu n +\bar {\Sigma^-}\gamma ^\mu \Sigma ^-
-\bar {\Sigma^+}\gamma ^\mu \Sigma ^+
- \bar {\Xi^-}\gamma ^\mu \Xi ^-
- 2 \bar {\Xi^0}\gamma ^\mu \Xi^0 \nonumber \\
&+& 3(3 \bar {\Delta^{-}} \gamma^\mu \Delta^{-} + \bar {\Delta^{+}} \gamma^\mu \Delta^{+} +2 \bar {\Delta^{0}} \gamma^\mu \Delta^{0} - \bar {\Sigma^{*+}}\gamma ^\mu \Sigma ^{*+} + \bar {\Sigma^{*-}}\gamma ^\mu \Sigma ^{*-}
\nonumber\\
& -&  \bar {\Xi^{*-}}\gamma ^\mu \Xi ^{*-} -2 \bar {\Xi^{*0}}\gamma ^\mu \Xi ^{*0} -3\bar {\Omega^-}\gamma ^\mu \Omega ^-)\Big)
\nonumber \\
& \times &
\Big(\bar {K^0} (\partial_\mu K^0) - (\partial_\mu {\bar {K^0}})  K^0 \Big )
\Big ] \nonumber \\
& + & \frac{m_K^2}{2f_K} \left[\left(\sigma + \sqrt{2} \zeta + \delta\right) K^+ K^- +
 \left(\sigma + \sqrt{2} \zeta - \delta\right) K^0 \bar{K^0} \right] \nonumber \\
 &-& \frac{1}{f_K} \left[\left(\sigma + \sqrt{2} \zeta + \delta\right) (\partial_\mu K^+) (\partial^\mu K^-) +
  \left(\sigma + \sqrt{2} \zeta - \delta\right) (\partial_\mu K^0) (\partial^\mu \bar{K^0}) \right] \nonumber \\ 
 &+& \frac {d_1}{2 f_K^2} \Big[(\bar p p +\bar n n +\bar {\Lambda^0}{\Lambda^0}
+\bar {\Sigma ^+}{\Sigma ^+}
+\bar {\Sigma ^0}{\Sigma ^0}
+\bar {\Sigma ^-}{\Sigma ^-}
+\bar {\Xi ^-}{\Xi ^-}
+\bar {\Xi ^0}{\Xi ^0}
+ 3 (\bar {\Delta^{++}}\Delta^{++}
\nonumber\\
& + & \bar {\Delta^{+}} \Delta^{+} + \bar {\Delta^{-}} \Delta^{-} + \bar {\Delta^{0}} \Delta^{0} + \bar {\Sigma ^{*+}}{\Sigma ^{*+}}
+\bar {\Sigma ^{*0}}{\Sigma ^{*0}}
+\bar {\Sigma ^{*-}}{\Sigma ^{*-}}
+\bar {\Xi ^{*-}}{\Xi ^{*-}}
+\bar {\Xi ^{*0}}{\Xi ^{*0}} 
+\bar {\Omega ^{-}}{\Omega ^{-}})
 )\nonumber \\
&\times & \big ( (\partial _\mu {K^+})(\partial ^\mu {K^-})
+(\partial _\mu {K^0})(\partial ^\mu {\bar {K^0}})
\big )\Big] \nonumber \\
 &+& \frac {d_2}{2 f_K^2} \Big [
(\bar p p+\frac {5}{6} \bar {\Lambda^0}{\Lambda^0}
+\frac {1}{2} \bar {\Sigma^0}{\Sigma^0}
+\bar {\Sigma^+}{\Sigma^+}
+\bar {\Xi^-}{\Xi^-}
+\bar {\Xi^0}{\Xi^0}
+3(\bar {\Delta^{++}} \Delta^{++} + \frac {2}{3}  \bar {\Delta^{+}} \Delta^{+}
\nonumber \\
& + & \frac {1}{3} \bar {\Delta^{0}} \Delta^{0} + \bar {\Sigma ^{*+}}{\Sigma ^{*+}}
+ \frac {2}{3} \bar {\Sigma ^{*0}}{\Sigma ^{*0}}
+ \frac {1}{3} \bar {\Sigma ^{*-}}{\Sigma ^{*-}}
+ \frac {2}{3} \bar {\Xi ^{*-}}{\Xi ^{*-}}
+\bar {\Xi ^{*0}}{\Xi ^{*0}} 
+\bar {\Omega ^{-}}{\Omega ^{-}}
)) (\partial_\mu K^+)(\partial^\mu K^-) 
\nonumber \\
 &+ &(\bar n n
+\frac {5}{6} \bar {\Lambda^0}{\Lambda^0}
+\frac {1}{2} \bar {\Sigma^0}{\Sigma^0}
+\bar {\Sigma^-}{\Sigma^-}
+\bar {\Xi^-}{\Xi^-}
+\bar {\Xi^0}{\Xi^0}
+ 3(\frac {1}{3}  \bar {\Delta^{+}} \Delta^{+} + \frac {2}{3}  \bar {\Delta^{0}} \Delta^{0} +\bar {\Delta^{-}} \Delta^{-}
\nonumber \\
&+& \frac {1}{3}  \bar {\Sigma ^{*+}}{\Sigma ^{*+}}
+ \frac {2}{3} \bar {\Sigma ^{*0}}{\Sigma ^{*0}}
+ \bar {\Sigma ^{*-}}{\Sigma ^{*-}}
+\frac {2}{3} \bar {\Xi ^{*0}}{\Xi ^{*0}}
+ \bar {\Xi ^{*-}}{\Xi ^{*-}}
+{\bar {\Omega ^{-}}}{\Omega ^{-}}
)) (\partial_\mu K^0)(\partial^\mu {\bar {K^0}})
\Big ]. 
\end{eqnarray}
\par
In the above, the parameters $d_1$ and $d_2$ are assigned values of $ 2.56/m_K $ and $ 0.73/m_K$, respectively \cite{Mishra:2008kg,Mishra:2008dj}. These values were determined by fitting to experimental measurements of the kaon-nucleon scattering length \cite{barnes1994kaon}.
Using the Euler -Lagrangian equation of motion and applying Fourier transformation, the dispersion relations are obtained from
 Eq.~(\ref{Eq_kBD_gen2}) to calculate the medium-modified masses of K ($K^+$, $K^0$) and $\bar K$ ($K^-$, $\bar {K^0}$) mesons. Therefore, we have  
\begin{equation}
-\omega^2+ {\vec k}^2 + m_{K (\bar K)}^2 -\Pi^*(\omega, |\vec k|)=0,
\label{eq_self_energy}
\end{equation}
where the in-medium self-energy for $K$ and $\bar K$ mesons is represented by $\Pi^*(\omega, |\vec k|)$.
The kaon isospin doublet $K$ ($K^+$, $K^0$) has its self-energy expressed as
\begin{eqnarray}
\Pi^*_K (\omega, |\vec k|) &= & -\frac {1}{4 f_K^2}\Big [3 (\rho_p +\rho_n)
\pm (\rho_p -\rho_n) \pm 2 (\rho_{\Sigma^+}-\rho_{\Sigma^-})
-\big ( 3 (\rho_{\Xi^-} +\rho_{\Xi^0}) \pm (\rho_{\Xi^-} 
\nonumber\\
&-&\rho_{\Xi^0})\big) +
3\Big\{\big(3(\rho_{\Delta^+} + \rho_{\Delta^0})\pm (\rho_{\Delta^+} - \rho_{\Delta^0}) \big) \pm 2 (\rho_{\Sigma^{*+}}- \rho_{\Sigma^{*-}}) \nonumber\\
&-& \big(3(\rho_{\Xi^{*-}} + \rho_{\Xi^{*0}})\pm (\rho_{\Xi^{*-}} - \rho_{\Xi^{*0}}) \big)
- 6 \rho_{\Omega^-} +6 \big( a \rho_{\Delta^{++}} + b \rho_{\Delta^-})   \big)\Big\} \Big ] \omega \nonumber\\
&+& \frac {m_K^2}{2 f_K} (\sigma ' +\sqrt 2 \zeta ' \pm \delta ')
\nonumber \\ & +& \Big [- \frac {1}{f_K}
(\sigma ' +\sqrt 2 \zeta ' \pm \delta ')
+\frac {d_1}{2 f_K ^2} \Big ({\rho^s}_{p} +{\rho^s}_{n}
+{\rho^s} _{\Lambda^0}+ {\rho^s} _{\Sigma^+}+{\rho^s} _{\Sigma^0}
+{\rho^s} _{\Sigma^-} \nonumber\\
&+&{\rho^s} _{\Xi^-} +{\rho^s} _{\Xi^0}+ 3\Big \{{\rho^s}_{\Delta^{++}} + {\rho^s}_{\Delta^+}+ {\rho^s}_{\Delta^0} + {\rho^s}_{\Delta^-} + {\rho^s} _{\Sigma^{*+}}+{\rho^s} _{\Sigma^{*0}} + {\rho^s} _{\Sigma^{*-}} \nonumber\\
&+& {\rho^s} _{\Xi^{*-}} +{\rho^s} _{\Xi^{*0}} + {\rho^s}_{\Omega^-} \Big  \}\Big)
+\frac {d_2}{4 f_K ^2} \Big (({\rho^s} _p +{\rho^s} _n)
\pm   ({\rho^s} _p -{\rho^s} _n)
+{\rho^s} _{\Sigma ^0}+\frac {5}{3} {\rho^s} _{\Lambda^0} \nonumber\\
&+& ({\rho^s} _{\Sigma ^+} + {\rho^s} _{\Sigma ^-})
\pm ({\rho^s} _{\Sigma ^+}-{\rho^s} _{\Sigma ^-})
 +  2 {\rho^s} _ {\Xi^-} + 2 {\rho^s} _ {\Xi^0} + 3 \Big\{ 2 ( a \rho^s_{\Delta^{++}} + b \rho^s_{\Delta^-}) \nonumber\\ 
 &+& \frac{1}{3} (3(\rho^s_{\Delta^+} + \rho^s_{\Delta^0})\pm (\rho^s_{\Delta^+} - \rho^s_{\Delta^0})) + \frac{4}{3} \rho^s_{\Sigma^{*0}} + \frac{1}{3} (4(\rho^s_{\Sigma^{*+}} 
 + \rho^s_{\Sigma^{*-}})\pm 2 (\rho^s_{\Sigma^{*+}} - \rho^s_{\Sigma^{*-}})) \nonumber\\
 &+& \frac{1}{3} (5(\rho^s_{\Xi^{*0}} + \rho^s_{\Xi^{*-}})\pm (\rho^s_{\Xi^{*0}} - \rho^s_{\Xi^{*-}})) +2 \rho^s_{\Omega^-}\Big\}
\Big )
\Big ]
(\omega ^2 - {\vec k}^2).
\label{sek}
\end{eqnarray}
The symbols $\sigma '$, $\zeta '$, and $\delta '$ represent the variations of the $\sigma$, $\zeta$, and $\delta$ fields from their corresponding vacuum expectation values, which are symbolized as $\sigma_{0}$, $\zeta_{0}$, and $\delta_{0}$. The following expression is derived for the self energy of antikaon isospin doublet $\bar{K}$ ($K^-$, $\bar {K^0}$)
\begin{eqnarray}
\Pi^*_{\bar K} (\omega, |\vec k|) &= & \frac {1}{4 f_K^2}\Big [3 (\rho_p +\rho_n)
\pm (\rho_p -\rho_n) \pm 2 (\rho_{\Sigma^+}-\rho_{\Sigma^-})
-\big ( 3 (\rho_{\Xi^-} +\rho_{\Xi^0}) \pm (\rho_{\Xi^-} 
\nonumber\\
&-&\rho_{\Xi^0})\big) +
3\Big\{\big(3(\rho_{\Delta^+} + \rho_{\Delta^0})\pm (\rho_{\Delta^+} - \rho_{\Delta^0}) \big) \pm 2 (\rho_{\Sigma^{*+}}- \rho_{\Sigma^{*-}}) \nonumber\\
&-& \big(3(\rho_{\Xi^{*-}} + \rho_{\Xi^{*0}})\pm (\rho_{\Xi^{*-}} - \rho_{\Xi^{*0}}) \big)
- 6 \rho_{\Omega^-} +6 \big( a \rho_{\Delta^{++}} + b \rho_{\Delta^-})   \big)\Big\} \Big ] \omega \nonumber\\
&+& \frac {m_K^2}{2 f_K} (\sigma ' +\sqrt 2 \zeta ' \pm \delta ')
\nonumber \\ & +& \Big [- \frac {1}{f_K}
(\sigma ' +\sqrt 2 \zeta ' \pm \delta ')
+\frac {d_1}{2 f_K ^2} \Big ({\rho^s}_{p} +{\rho^s}_{n}
+{\rho^s} _{\Lambda^0}+ {\rho^s} _{\Sigma^+}+{\rho^s} _{\Sigma^0}
+{\rho^s} _{\Sigma^-} \nonumber\\
&+&{\rho^s} _{\Xi^-} +{\rho^s} _{\Xi^0}+ 3\Big \{{\rho^s}_{\Delta^{++}} + {\rho^s}_{\Delta^+}+ {\rho^s}_{\Delta^0} + {\rho^s}_{\Delta^-} + {\rho^s} _{\Sigma^{*+}}+{\rho^s} _{\Sigma^{*0}} + {\rho^s} _{\Sigma^{*-}} \nonumber\\
&+& {\rho^s} _{\Xi^{*-}} +{\rho^s} _{\Xi^{*0}} + {\rho^s}_{\Omega^-} \Big  \}\Big)
+\frac {d_2}{4 f_K ^2} \Big (({\rho^s} _p +{\rho^s} _n)
\pm   ({\rho^s} _p -{\rho^s} _n)
+{\rho^s} _{\Sigma ^0}+\frac {5}{3} {\rho^s} _{\Lambda^0} \nonumber\\
&+& ({\rho^s} _{\Sigma ^+} + {\rho^s} _{\Sigma ^-})
\pm ({\rho^s} _{\Sigma ^+}-{\rho^s} _{\Sigma ^-})
 +  2 {\rho^s} _ {\Xi^-} + 2 {\rho^s} _ {\Xi^0} + 3 \Big\{ 2 ( a \rho^s_{\Delta^{++}} + b \rho^s_{\Delta^-}) \nonumber\\ 
 &+& \frac{1}{3} (3(\rho^s_{\Delta^+} + \rho^s_{\Delta^0})\pm (\rho^s_{\Delta^+} - \rho^s_{\Delta^0})) + \frac{4}{3} \rho^s_{\Sigma^{*0}} + \frac{1}{3} (4(\rho^s_{\Sigma^{*+}} 
 + \rho^s_{\Sigma^{*-}})\pm 2 (\rho^s_{\Sigma^{*+}} - \rho^s_{\Sigma^{*-}})) \nonumber\\
 &+& \frac{1}{3} (5(\rho^s_{\Xi^{*0}} + \rho^s_{\Xi^{*-}})\pm (\rho^s_{\Xi^{*0}} - \rho^s_{\Xi^{*-}})) +2 \rho^s_{\Omega^-}\Big\}
\Big)
\Big ]
(\omega ^2 - {\vec k}^2).
\label{se_antik}
\end{eqnarray}
In the previously mentioned equations, the ± sign gives $\Pi^*_{ K^+}$ ($\Pi^*_{ K^-}$) and $\Pi^*_{K^{0}}$ ($\Pi^*_{\bar K^{0}}$) respectively. The parameters are assigned as $a=1$, $b=0$ for $K^{+}$ and $K^{-}$, but $a=0$ and $b=1$ are used for $K^{0}$ and $\bar K^{0}$. 
The in-medium masses of $K (\bar K)$ mesons in the dense resonance matter are determined at momentum $|\vec k|=0$ using Eq.~(\ref{eq_self_energy}) with $m_K$ and $m_{\bar K}$  denote the vacuum masses of kaons and antikaons, respectively.

\subsection{\label{subsec2.3} In-medium mass and decay width of $\phi$ meson}
\label{SEC_IIC}
\begin{figure}[h]
\includegraphics[scale=0.1]{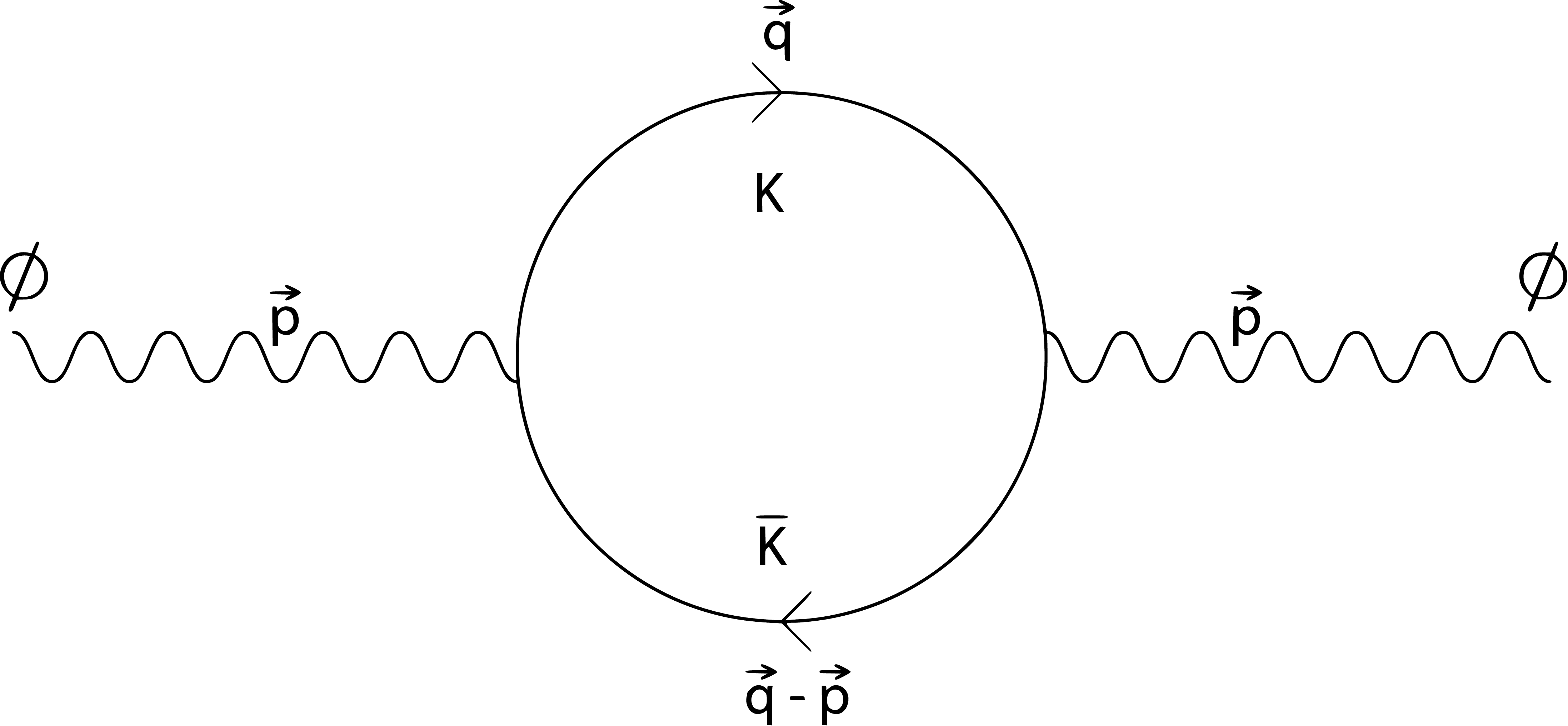}
\caption{ $\phi K \bar{K}$ interaction at one loop level.}
\label{loop}
\end{figure}
The in-medium masses of $K$ and $\bar K$ mesons  calculated in the isospin asymmetric resonance medium are used as input in the current subsection to compute the in-medium properties of $\phi$ for the decay channel $\phi$ $\rightarrow$ $K \bar K$ at one loop level as depicted in Fig. \ref{loop}.   
The interaction Lagrangian density for coupling of $\phi$ meson to $K \bar K$ is given as \cite{Ko:1992tp,Klingl:1996by} 

\begin{equation}
\label{eqn:phikk}
\mathcal{L}_{\phi K \bar {K}} = i g_{\phi}\phi^{\mu}
\left[ \bar K(\partial_{\mu} K)-(\partial_{\mu} \bar K)K\right].
\end{equation}

The present study has not taken into account the $\phi\phi K \bar{K}$ interactions, as their impact on in-medium masses and decay width of $\phi$ meson is minimal compared to the $\phi K \bar K$ interactions \cite{Cobos-Martinez:2017vtr}. 
The scalar part of the in-medium self-energy corresponding to the loop diagram for $\phi$ meson at rest can be expressed as
\begin{equation}
\label{eqn:phise}
i\Pi^*_{\phi}(p)=-\frac{8}{3}g_{\phi}^{2}\int \frac {d^4q}{(2\pi)^4} \vec{q}^{\,2}
D_{K}(q)D_{\bar K}(q-p) \,. 
\end{equation}
In the above, $D_{K}(q)$=$\left(q^{2}-m_{K}^{*^{2}}+i\epsilon\right)^{-1}$ and $D_{\bar K}(q$-$p)$=$\left((q-p)^{2}-m_{\bar K}^{*^{2}}+i\epsilon\right)^{-1}$ represent the kaon and antikaon propagators, respectively. The $\phi$ meson four-momentum vector is denoted by $p=(p^{0}=m^*_{\phi},\vec{0})$, with $m^*_{\phi}$ indicating the in-medium mass of $\phi$ meson. 
The average masses of kaon and antikaon are symbolized by $m^*_{K}$ (=$\frac{m_{K^+}^{*}+m_{K^0}^{*}}{2}$) and $m^*_{\bar K}$(=$\frac{m_{K^-}^{*}+m_{\bar K^0}^{*}}{2}$), respectively.
Values of $m_{K^+}^{*}$, $m_{K^0}^{*}$, $m_{K^-}^{*}$ and $m_{\bar K^0}^{*}$ are calculated by using Eq.~(\ref{eq_self_energy}) for finite $T$, $I_a$, and $f_s$ within the medium.
The effective mass of the $\phi$ meson is calculated from the real part of $\Pi^*_{\phi}(p)$ through relation \cite{Cobos-Martinez:2017vtr}
\begin{equation}
\label{eqn:phimassvacuum}
m_{\phi}^{*^{2}}=\left(m_{\phi}^{0}\right)^{2}+ \text{Re}\Pi^*_{\phi}(m_{\phi}^{*^{2}}),
\end{equation}
where $m_{\phi}^{0}$ is the bare mass of the $\phi$ meson. The real part of self-energy is given as \cite{Cobos-Martinez:2017vtr, kumar2020phi}
\begin{equation}
\label{eqn_sephi}
\text{Re}\Pi^*_{\phi}=-\frac{4}{3}g_{\phi}^{2} \, \mathcal{P}\!\!
\int \frac {d^3q} {(2\pi)^3} \vec{q}^{\,2}\frac{(E^*_K+E^*_{\bar K})}{E^*_{K} E^*_{\bar K} ((E^*_K+E^*_{\bar K})^2-m_{\phi}^{*^2})} \, ,
\end{equation}
where $E^*_{K/ \bar K}=(\vec{q}^{\,2}+m_{K/ \bar K}^{*^2})^{1/2}$ and $\mathcal{P}$ stands for principal value of the integral. The integral in 
Eq.~(\ref{eqn_sephi}) exhibits divergence but must be regularized. To address this, we employ a phenomenological vertex form factor that incorporates a cutoff parameter $\Lambda_c$ as in Ref. \cite{Krein:2010vp}. 
The loop integral after renormalization is expressed as
\begin{equation}
\label{eqn:regphi}
\text{Re}\Pi^*_{\phi}=-\frac{4}{3}g_{\phi}^{2} \, \mathcal{P}\!\!
\int^{\Lambda_c}_{0}  \frac {d^3q} {(2\pi)^3} \vec{q}^{\,4}\left( \frac{\Lambda^2_c+m_{\phi}^{*^2}}{\Lambda^2_c+4E_{K}^{*^2}}\right)^4 \frac{(E^*_K+E^*_{\bar K})}{E^*_{K} E^*_{\bar K} ((E^*_K+E^*_{\bar K})^2-m_{\phi}^{*^2})} \, .
\end{equation}
By using a quark pair creation model and applying the Gaussian wave function of mesons, the vertex form factor can be calculated to get a rough approximation of $\Lambda_c$ \cite{Krein:2010vp, Close:2005se, Godfrey:1985xj, kumar2020phi}.
The study in Ref. \cite{Krein:2010vp} calculated the in-medium masses of $J/ \psi$ using vertex form factors, with $\Lambda_c$ values vary in the range $1 \text{ GeV} \leq \Lambda_c \leq 3 \text{ GeV}$. A comparable range for $\Lambda_c$ (1 to 4 GeV) was employed to determine the masses of $\phi$ meson within hadronic medium in Ref. \cite{Cobos-Martinez:2017vtr,kumar2020phi}. Thus, we consider values of $\Lambda_c$ in the range  2 to 4 GeV to understand its impact on the properties of the $\phi$ meson in isospin asymmetric resonance medium. 
 The decay width of $\phi$ meson is derived from the imaginary component of the self-energy Im$\Pi^*_{\phi}$ and given as \cite{Li:1994cj}   
 \begin{equation}
\label{eqn:phidecaywidth}
\Gamma^*_{\phi}  = \frac{g_{\phi}^{2}}{24\pi
} \frac{1}{m_{\phi}^{*^5}} 
\left((m_{\phi}^{*^2}-(m_{K}^{*}+m_{\bar K}^{*})^2)(m_{\phi}^{*^2}-(m_{K}^{*}-m_{\bar K}^{*})^2)\right)^{3/2} \, .
\end{equation}
In the vacuum, the empirical width of the $\phi$ meson is used to calculate the coupling constant $g_\phi$, which is 4.539 in this work.

\section{Results and discussion}
	\label{results}
In this section, we discuss the results of in-medium properties of $K$, $\bar K$, and $\phi$ mesons in dense resonance medium. 
\begin{table}
\centering
\def\arraystretch{1.8}
\begin{tabular}{cccccccccc}
\hline \hline
 Baryon &~$m_i$ (MeV)&${g_{\sigma i}}$& ${g_{\zeta i}}$  &   ${g_{\delta i}}$ &   ${g_{\omega i}}$ &${g_{\rho i}}$ &${g_{\phi i}}$  & $m^i_3$  &  $U_i$ (MeV)\\ \hline \hline
N&939&9.83 &  -1.22 &  2.34 &  11.8 &  4.7 &  0&-0.11&-71\\ 
$\Lambda$&1115&5.31 & 5.80 & 0 & 8.89  & 0 & -6.29& 0.44 & -28 \\ 
$\Sigma$&1193&6.13 & 5.80 & 6.79 & 8.89 & 8.89 & -6.28& 2.12 & 30 \\ 
$\Xi$&1315&3.68 & 9.14 & 2.36  & 4.44 & 4.44 & -12.56 & 1.6 & -18 \\
$\Delta$&1232&10.74 & 2.20 & 2.49 & 12.74 & 12.74 & 0& 0.68 & -80   \\ 
$\Sigma^*$&1385&8.20 & 5.80 & 3.47 & 8.50  & 8.50 & -6.01& 1.45 & -64 \\ 
$\Xi^*$&1530&5.66 & 9.40 & 2.34 & 4.25 & 4.25 & -12.01& 1.28 & -19 \\ 
$\Omega$ &1672&3.12 & 15.19 & 0  & 0 & 0 & -18.02& 3.44 & -25 \\ \hline \hline
\end{tabular}
\caption{Different coupling constants,  parameter $m^i_3$ and the baryonic potential $U_i$ used in present calculations.}
\label{coupling}
\end{table}
We mainly examine the impact of baryonic density $\rho_B$,  temperatures $T$, isospin asymmetry $I_a$, and strangeness fraction $f_s$ of resonance medium on the medium-modified mass and decay width of $\phi$ meson.
The coupling constants that reflect the strength of interaction of scalar and vector fields with baryons, as well as the baryonic potential, $U_i$, 
are presented in Table \ref{coupling}.
The parameter $m^i_3$ is fitted to reproduce the reasonable baryonic potential through the relation, $U_i= M^*_{i}- m_{i}+g_{\omega i} \omega+g_{\rho i}\tau_{3i} \rho+g_{\phi i} \phi$, 
where $M^*_{i}= m^*_{i}+ m^i_3 \left( \sqrt{2} (\sigma - \sigma_0) + (\zeta - \zeta_0) \right)$ 
at nuclear saturation density for symmetric nuclear matter. 
The other parameters of the chiral SU(3) hadronic mean-field model are specified as follows, $k_0=2.37$, $k_1=1.40$, $k_2=-5.55$, $k_3=-2.65$, $k_4=-0.23$, $g_4=38.9$, $d=0.06$,  $m_\pi=139$ MeV,  pion decay constant $f_\pi=93.3$ MeV, $m_K=498$ MeV, and the kaon decay constant $f_K$= 122 MeV. The vacuum values for the scalar fields are $\sigma_0 = -93.3$ MeV, $\zeta_0 = -106.76$ MeV, and $\chi_0 = 401.91$ MeV.   
 \subsection{The in-medium mass of kaon and antikaon in resonance matter}
 \label{results_kaon_mass}
\begin{figure}[hbt!]
 	\begin{subfigure}{.5\linewidth}
 		(a)\includegraphics[width=0.9\linewidth]{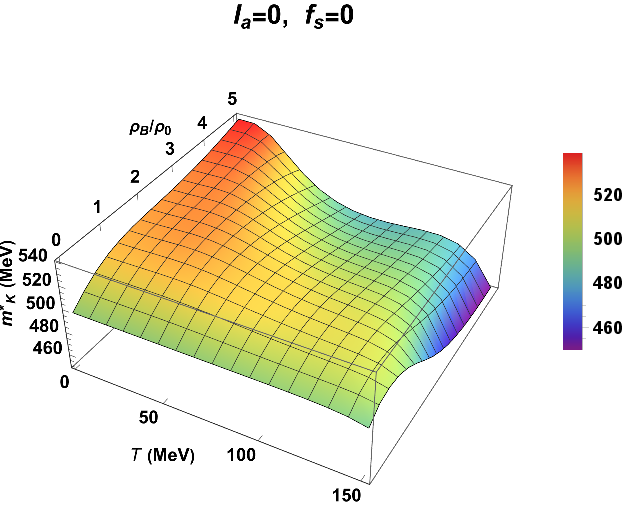}
 		
 	\end{subfigure}\hfill 
 	\begin{subfigure}{.5\linewidth}
 		(b)\includegraphics[width=0.9\linewidth]{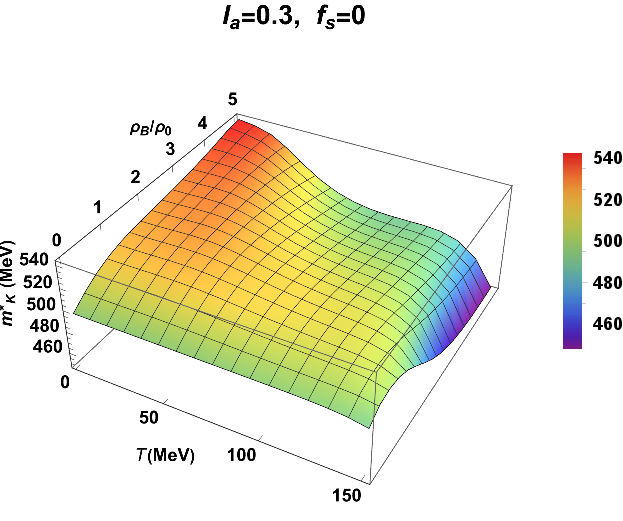}
 	\end{subfigure}
 	\begin{subfigure}{.5\linewidth}
 		(c)\includegraphics[width=0.9\linewidth]{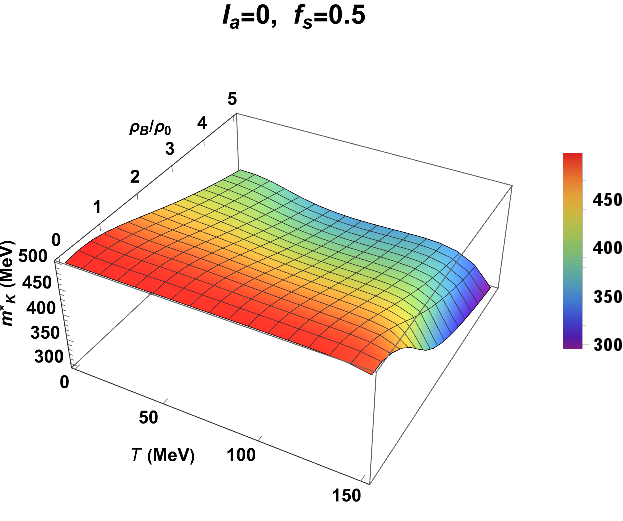}
 	\end{subfigure}\hfill 
 	\begin{subfigure}{.5\linewidth}
 		(d)\includegraphics[width=0.9\linewidth]{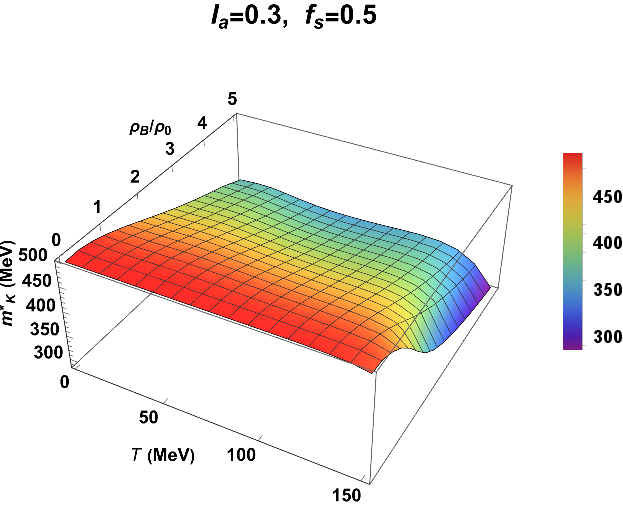}
 	\end{subfigure}
 	\caption{
 		The average effective kaon mass $m^*_{K}$ is plotted for resonance medium as a function of baryonic density ratio $\rho_B/\rho_0$ and temperature $T$ under various conditions of isospin asymmetry parameter $I_a$ and strangeness fraction $f_s$ of the medium. }
 	\label{kaonmass}
 \end{figure}

\begin{figure}[hbt!]
 	\begin{subfigure}{.5\linewidth}
 		(a)\includegraphics[width=0.9\linewidth]{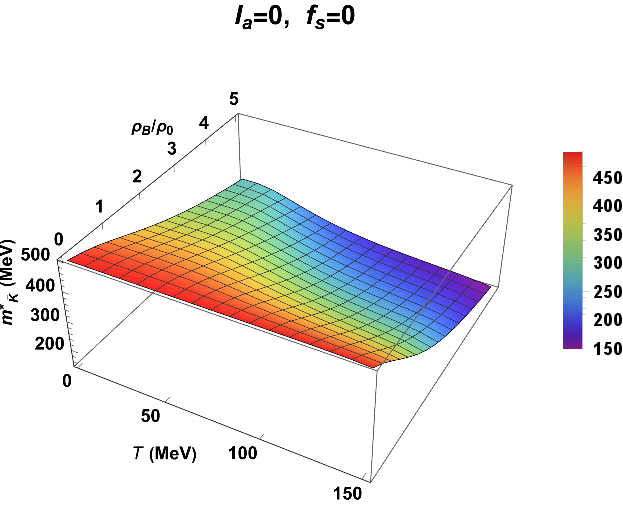}
 		
 	\end{subfigure}\hfill 
 	\begin{subfigure}{.5\linewidth}
 		(b)\includegraphics[width=0.9\linewidth]{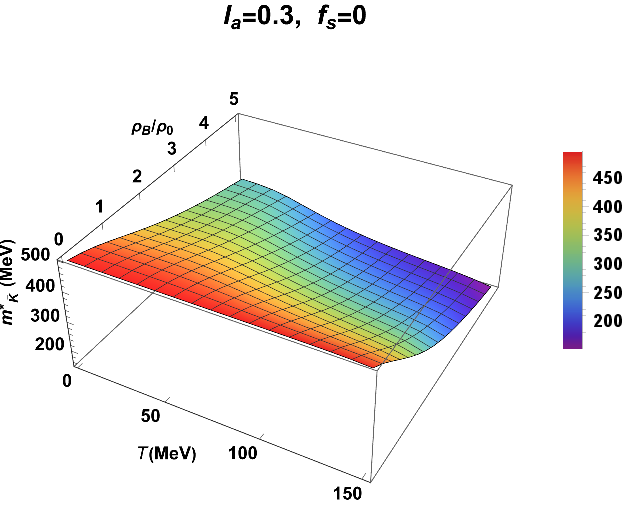}
 	\end{subfigure}
 	\begin{subfigure}{.5\linewidth}
 		(c)\includegraphics[width=0.9\linewidth]{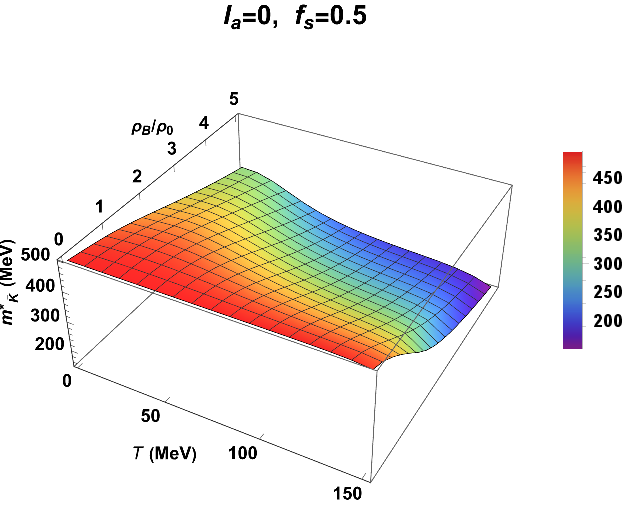}
 	\end{subfigure}\hfill 
 	\begin{subfigure}{.5\linewidth}
 		(d)\includegraphics[width=0.9\linewidth]{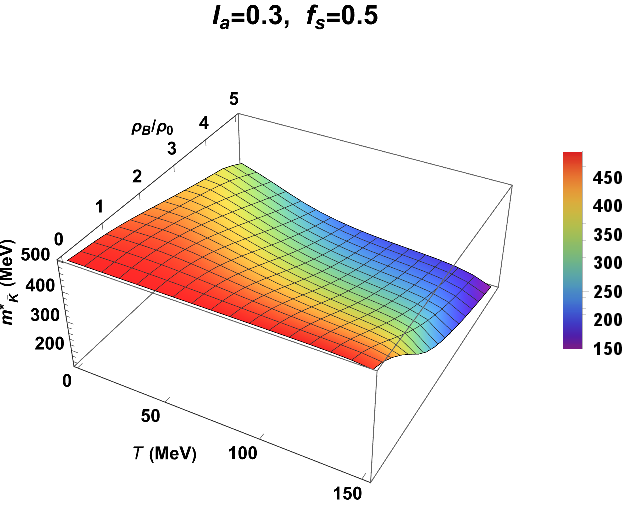}
 	\end{subfigure}
 	\caption{
 		The average effective antikaon mass $m^*_{\bar K}$ is plotted for resonance medium with respect to the baryonic density ratio $\rho_B/\rho_0$ and temperature $T$.
        The impact of isospin asymmetry parameter $I_a$ and strangeness fraction $f_s$ of the medium on $m^*_{\bar K}$ is also considered.}
        \label{antikaonmass}
 \end{figure}

 In this subsection, we discuss the results of medium-modified kaon and antikaon masses in isospin asymmetric resonance matter at finite density and temperature.  In order to compute the in-medium properties of $\phi$ meson, it is essential first to determine in-medium $K$ ($\bar K$)  masses as detailed in Ref. \cite{Cobos-Martinez:2017vtr, kumar2020phi}. 
 The chiral SU(3) model is employed to calculate effective $K$ ($\bar K$) masses by utilizing medium-modified scalar fields ($\sigma$, $\zeta$, and $\delta$) along with vector and scalar baryonic densities, as described in self-energy expressions given in Eqs. (\ref{sek}) and (\ref{se_antik}).
Our previous study \cite{Kaur:2024cfm} discussed how the presence of resonance baryons in the medium leads to significant modifications in the effective masses of $K$ ($K^+$, $K^0$) and $\bar K$ ($\bar K^0$, $\bar K^0$). 
In the present study, the average kaon mass, $m^*_{K}=\frac{m^*_{K^{+}}+m^*_{K^0}}{2}$, and antikaon mass, $m^*_{\bar K}=\frac{m^*_{K^{0}}+m^*_{\bar K^0}}{2}$, represented in  Figs.\ref{kaonmass} and \ref{antikaonmass}, are utilized to determined the in-medium mass of $\phi$ meson. These figures illustrate the 3D representation of the effective mass of kaon ($m^*_{K}$) and antikaon ($m^*_{\bar K}$) plotted with respect to baryonic density ratio $\rho_B/\rho_0$  and temperature $T$ in dense resonance matter for isospin asymmetry $I_a = 0$ and $0.3$, and strangeness fractions $f_s = 0$ and $0.5$.  
 \par
 In Fig. \ref{kaonmass} (a), we consider a symmetric medium ($I_a=0$), comprising of non-strange particles ($p, n$, and $\Delta^{++,+,0,-}$), i.e., $f_s=0$ case. 
At lower temperature, the effective kaon mass is observed to increase as baryonic density rises from zero to $5\rho_0$, while at higher temperature, the opposite trend is observed with an increase in $\rho_B/ \rho_0$ as exhibited in Fig. \ref{kaonmass}(a). 
The effective masses of $K$ and $\bar{K}$ mesons in the chiral SU(3) model are evaluated by incorporating the effects of various interaction terms into the total Lagrangian density, as specified in Eq. (\ref{Eq_kBD_gen1}). Among the interaction terms in this Lagrangian, the vectorial interaction term has a significant impact on $m^*_{K}$, causing a repulsive influence at low temperatures and an attractive effect at high temperatures. In Fig. \ref{kaonmass}(b), for asymmetric medium with $I_a=0.3$, where a comparable trend is observed as in Fig. \ref{kaonmass} (a), but the change in magnitude of $m^*_{K}$ from its vacuum value (496 MeV) as a function of $\rho_B$ and $T$ increases slightly. The incorporation of strange particles within the medium at a given temperature, where all spin$-\frac{1}{2}$ octet $(p,n,\Lambda,\Sigma^{\pm,0}, \Xi^{-,0})$ as well as  spin$-\frac{3}{2}$ decuplet
($\Delta^{++,+,0,-}, \Sigma^{*\pm,0},\Xi^{*0,-}, \Omega^{-}$) baryons are considered, leads to a decrease in the effective mass of ${K}$ mesons due to dominant attractive interactions between kaons and baryons, as represented in Figs. \ref{kaonmass} (c) and (d). 
As can be seen from these results, the $m^*_{K}$ decreases with respect to baryonic density at lower temperature. However, at higher temperature, the rate of drop of kaon mass is more noticeable as shown in Figs. \ref{kaonmass} (c) and (d).
Similar to kaon masses, antikaon masses are plotted in Fig. \ref{antikaonmass}, and their respective results show the reduction in $m^*_{\bar K}$ with variation in temperature and baryonic density, regardless of values of $I_a$ and $f_s$. A comparison of Figs. \ref{kaonmass} and \ref{antikaonmass} reveal a more sizable reduction in antikaon masses compared to kaons at higher temperatures and densities. \par

\subsection{The in-medium mass of $\phi$ meson in resonance medium}
\begin{figure}
	\centering
	\includegraphics[width=0.9\linewidth]{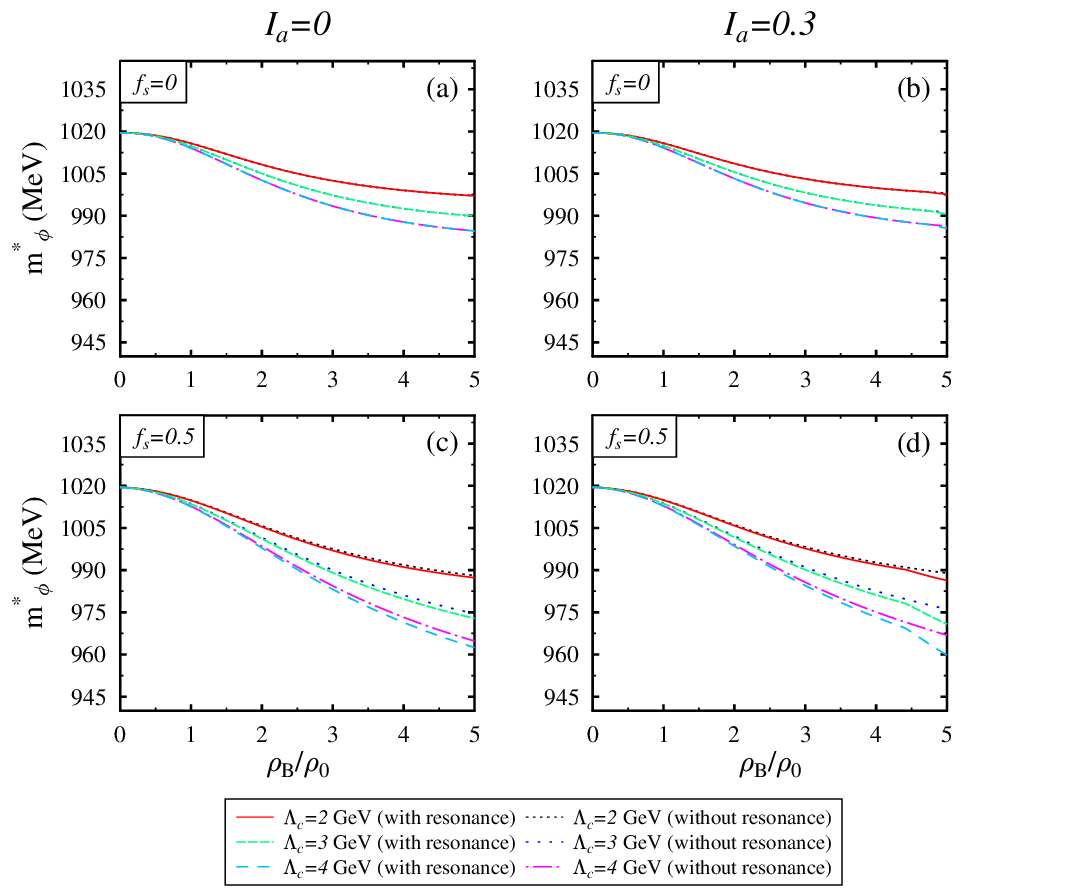}
	\caption{ Comparison of $m^*_\phi$ in resonance and non-resonance matter as a function of baryon density ratio $\rho_B/\rho_0$ for different values of $\Lambda_c$, $I_a$, and $f_s$ at temperature $T=0$ MeV.}
    
	\label{massT0}
\end{figure}

\begin{figure}
	\centering
	\includegraphics[width=0.9\linewidth]{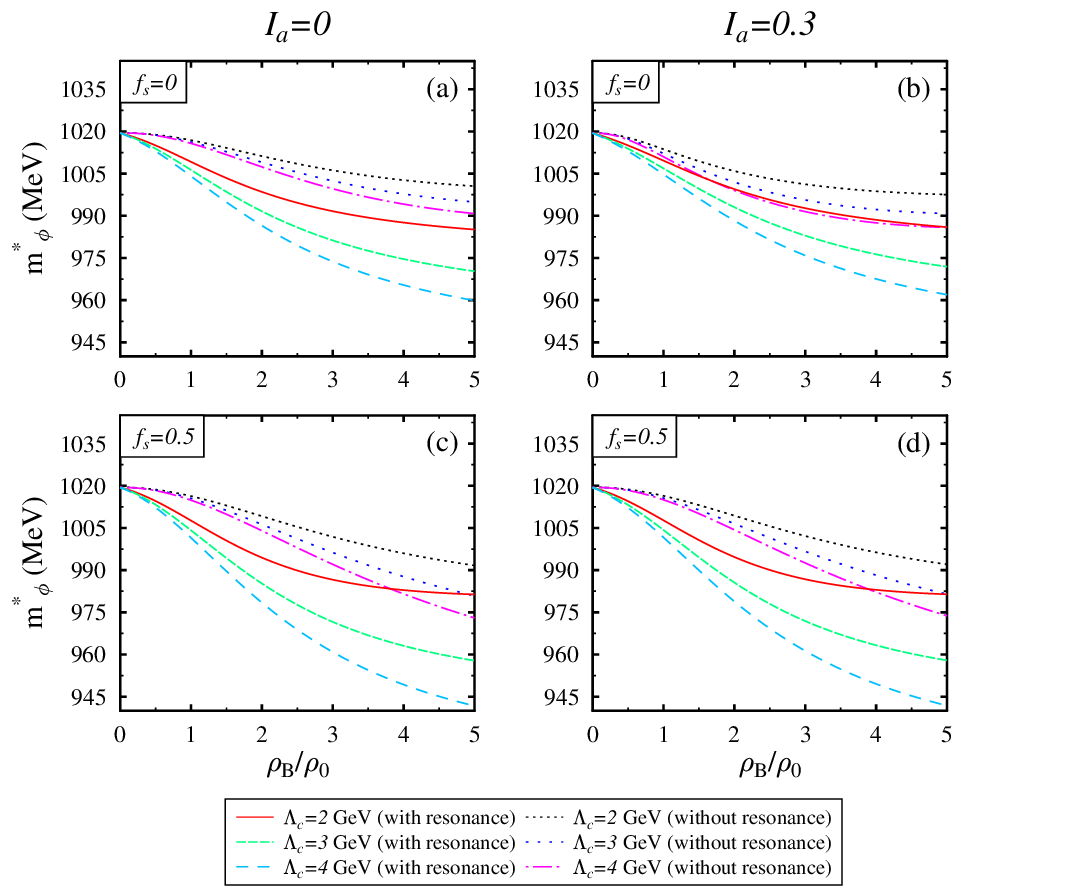}
	\caption{The comparison of $m^*_\phi$ in both resonance and non-resonance matter is presented as a function of baryon density ratio $\rho_B/\rho_0$, considering various values of $\Lambda_c$, $I_a$, and $f_s$ at temperature $T=100$ MeV.}
	\label{massT100}
\end{figure}

Next, we extend our numerical analysis to the in-medium $\phi$ meson properties in isospin asymmetric resonance matter. In Ref. \cite{Cobos-Martinez:2017vtr,Cobos-Martinez:2017woo}, the $\phi$ meson self energy in cold nuclear matter is caculated by employing quark-meson coulping model under simplifying assumption $m^{*}_{K}$=$m^{*}_{\bar K}$. However, the Ref \cite{chahal2024phi,kumar2020phi} and the current study treat these meson masses differently, i.e., $m^{*}_{K} \neq m^{*}_{\bar K}$ at finite temperature. To determine the medium-modified $\phi$ meson mass within the resonance medium, we plugged the findings of kaon and antikaon energies into Eq. (\ref{eqn_sephi}), the real part of the self-energy of the $\phi$ meson. 
\begin{table}
	\centering
	\small
	\begin{tabular}{|c|c|c|c|c|c|c|c|c|c|c|c|c|c|}
		\hline
		& & \multicolumn{4}{c|}{T=0 MeV}  & \multicolumn{4}{c|}{T=100 MeV}  & \multicolumn{4}{c|}{T=150 MeV} \\
		\cline{3-14}
		$m^*_{\phi}$ (MeV) &$f_s$ & \multicolumn{2}{c|}{$I_a=0$} & \multicolumn{2}{c|}{$I_a=0.3$} & \multicolumn{2}{c|}{$I_a=0$} & \multicolumn{2}{c|}{$I_a=0.3$} & \multicolumn{2}{c|}{$I_a=0$} & \multicolumn{2}{c|}{$I_a=0.3$} \\
		\cline{3-14}
				& &$~\rho_0~$&$~4\rho_0~$ &$~\rho_0~$&$~4\rho_0~$ &$~\rho_0~$&$~4\rho_0~$&$~\rho_0~$&$~4\rho_0~$
		&$~\rho_0~$&$~4\rho_0~$ &$~\rho_0~$&$~4\rho_0~$ \\ \hline
	$\Lambda_c=2$ GeV & 0 & 1015.7 & 999.1 & 1015.8 & 999.9 & 1009.2 & 987.6 & 1009.6 & 988.6 & 1002.9 & 983.8 & 1002.9 & 983.8\\
\cline{2-14}
 & 0.5 & 1014.8 & 991.1 & 1014.9 & 992.0 & 1007.6 & 982.9 & 1007.7 & 983.0 & 999.4 & 981.1 & 999.0 & 981.1\\
\cline{1-14}
$\Lambda_c=3$ GeV & 0 & 1014.8 & 992.6 & 1014.9 & 993.8 & 1006.3 & 974.6 & 1006.9 & 976.3 & 997.8 & 966.8 & 997.7 & 966.8\\
\cline{2-14}
 & 0.5 & 1013.5 & 979.7 & 1013.7 & 981.2 & 1004.1 & 963.1 & 1004.3 & 963.3 & 992.8 & 954.3 & 992.1 & 953.6\\
\cline{1-14}
$\Lambda_c=4$ GeV & 0 & 1014.1 & 987.8 & 1014.2 & 989.3 & 1004.1 & 965.4 & 1004.8 & 967.6 & 993.9 & 955.0 & 993.8 & 954.9\\
\cline{2-14}
 & 0.5 & 1012.6 & 971.4 & 1012.7 & 973.3 & 1001.5 & 949.3 & 1001.6 & 949.5 & 987.7 & 936.2 & 986.9 & 935.1\\
\cline{1-14}
\end{tabular}
	
	\caption{The in-medium values of $m^{*}_{\phi}$ at different fixed values of $\Lambda_c$, $\rho_B$, $I_a$, $f_s$, and $T$. }
	\label{table_mass}
\end{table}  
In Figs.~\ref{massT0} and \ref{massT100}, the  behavior of $\phi$ meson mass in isospin asymmetric dense resonance matter is analyzed by plotting it with respect to the baryon density ratio $\rho_B/\rho_0$ (in units of nuclear saturation density $\rho_0$) for different values of cutoff mass parameter $\Lambda_c$, at temperatures $T=0$ and 100 MeV, respectively. In each subplot, our results are presented for the scenarios that consider calculations with and without including decuplet baryons. The values of effective masses of $\phi$ mesons are tabulated at specific values of $\Lambda_c$, $\rho_B$, $I_a$, $f_s$, and $T$ in Table \ref{table_mass}. 
In Fig. \ref{massT0} (a) and (b), at $f_s=0$, considering nucleons and $\Delta$ baryons, the magnitude of $m^*_\phi$ decreases significantly with increasing the value of $\Lambda_c$ for both symmetric ($I_a =0$) and asymmetric ($I_a =0.3$) resonance medium. However, the reduction in $m^*_\phi$ with respect to $\Lambda_c$ is slightly less noticeable in the asymmetric matter across all $\rho_B$ values. We also observed that at $T=0$, the effective mass of $\phi$ meson almost remains the same for both pure nuclear matter and $\Delta$ resonance matter, i.e., regardless of the presence or absence of resonance decuplets, up to the 
considered $\rho_B/\rho_0$ ratio in this study, as depicted in Figs. \ref{massT0} (a) and (b). This consistency is attributed to the fact that $\Delta$ baryons are populated at higher baryonic densities, as elaborated in our previous investigation \cite{Kaur:2024cfm}. 

\begin{figure}[hbt!]
	\begin{subfigure}{.5\linewidth}
		(a)\includegraphics[width=0.9\linewidth]{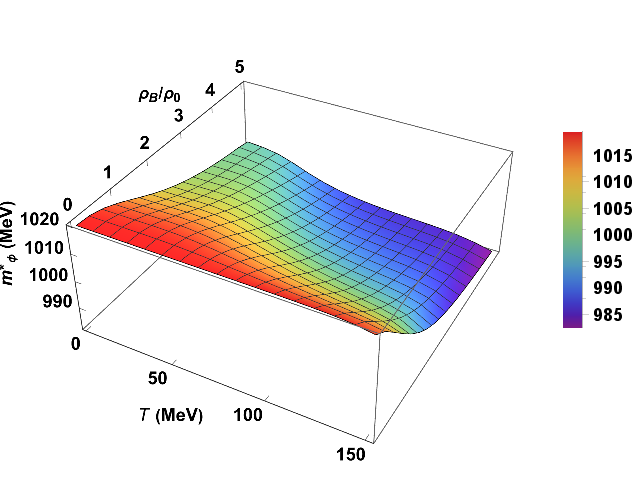}
		
	\end{subfigure}\hfill 
	\begin{subfigure}{.5\linewidth}
		(b)\includegraphics[width=0.9\linewidth]{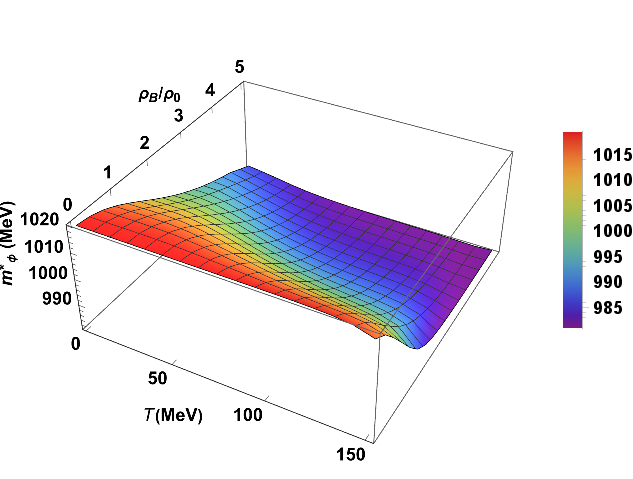}
	\end{subfigure}
	\begin{subfigure}{.5\linewidth}
		(c)\includegraphics[width=0.9\linewidth]{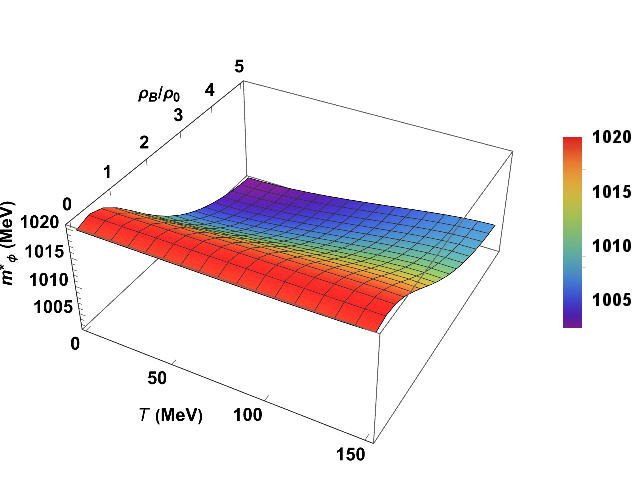}
	\end{subfigure}\hfill 
	\begin{subfigure}{.5\linewidth}
		(d)\includegraphics[width=0.9\linewidth]{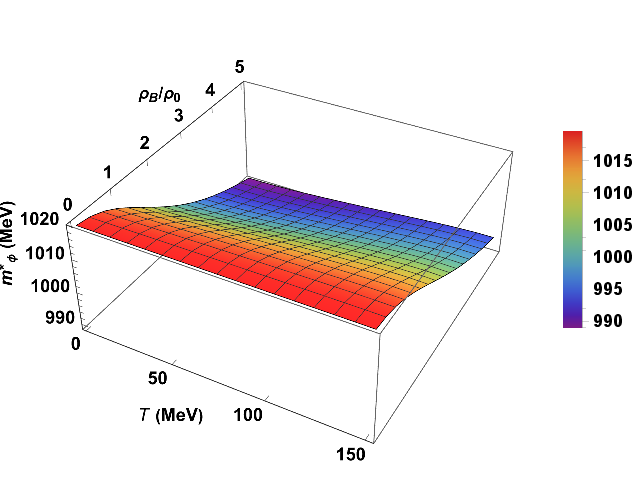}
	\end{subfigure}
	\caption{
		The three-dimensional representation of effective masses of $\phi$ meson is plotted as a function of baryon density ratio $\rho_B/\rho_0$ and temperature $T$ for isospin asymmetry $I_a=0.3$ and  cutoff parameter $\Lambda_c=2$ GeV. The left panel corresponds to $f_s = 0$, with (a) includes both nucleons and $\Delta$ baryons, while (c) considers only nucleons. The right panel is for $f_s=0.5$, where (b) incorporates all octet and decuplet baryons and (d) involves octet baryons only.}
	\label{3D_mphi_2CF}
\end{figure}
\begin{figure}[hbt!]
	\begin{subfigure}{.5\linewidth}
		(a)\includegraphics[width=0.9\linewidth]{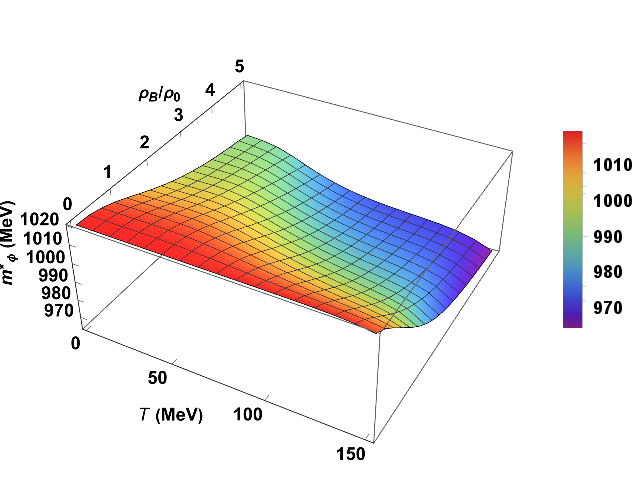}
		
	\end{subfigure}\hfill 
	\begin{subfigure}{.5\linewidth}
		(b)\includegraphics[width=0.9\linewidth]{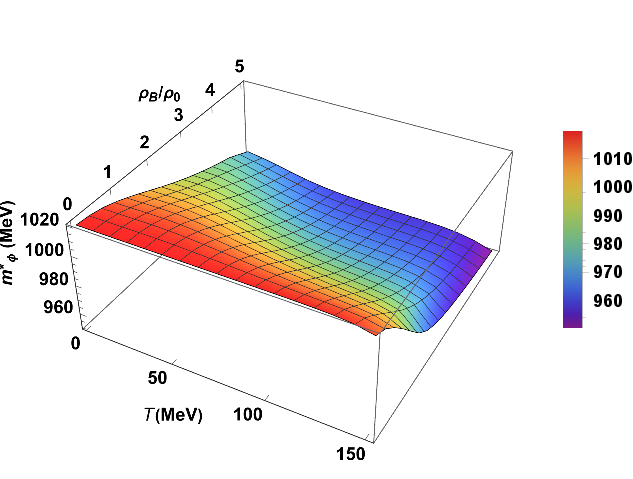}
	\end{subfigure}
	\begin{subfigure}{.5\linewidth}
		(c)\includegraphics[width=0.9\linewidth]{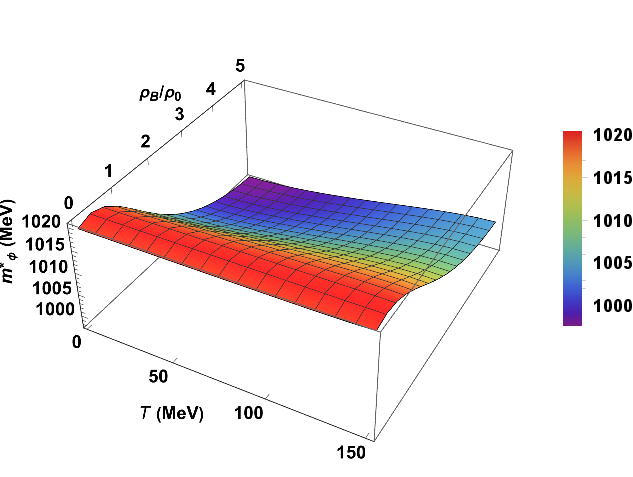}
	\end{subfigure}\hfill 
	\begin{subfigure}{.5\linewidth}
		(d)\includegraphics[width=0.9\linewidth]{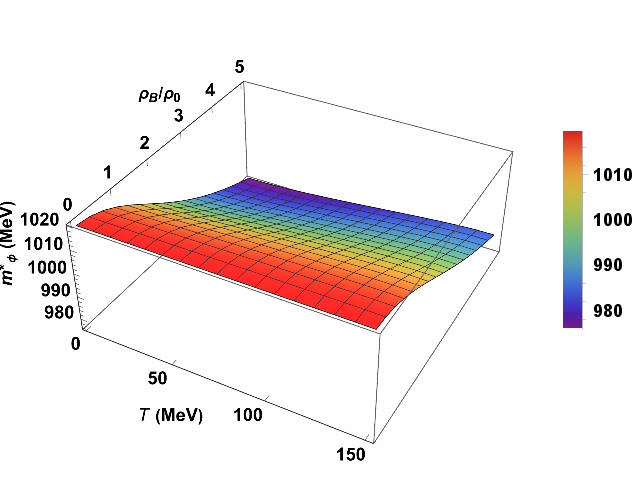}
	\end{subfigure}
	\caption{Same as Fig. \ref {3D_mphi_2CF}, at $\Lambda_c=3$ GeV.}
	\label{3D_mphi_3CF}
\end{figure}
\begin{figure}[hbt!]
	\begin{subfigure}{.5\linewidth}
		(a)\includegraphics[width=0.9\linewidth]{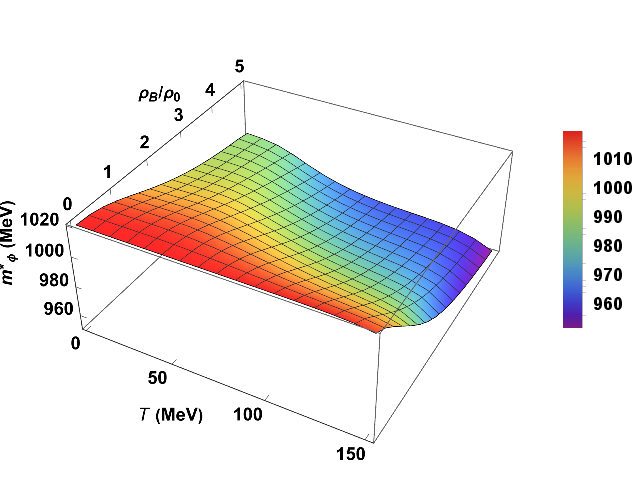}
		
	\end{subfigure}\hfill 
	\begin{subfigure}{.5\linewidth}
		(b)\includegraphics[width=0.9\linewidth]{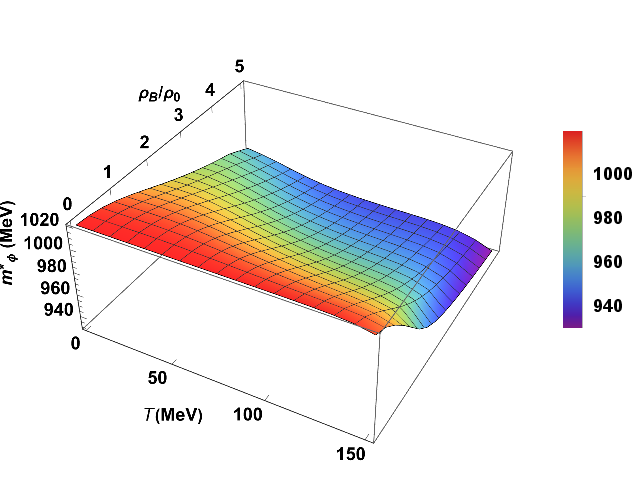}
	\end{subfigure}
	\begin{subfigure}{.5\linewidth}
		(c)\includegraphics[width=0.9\linewidth]{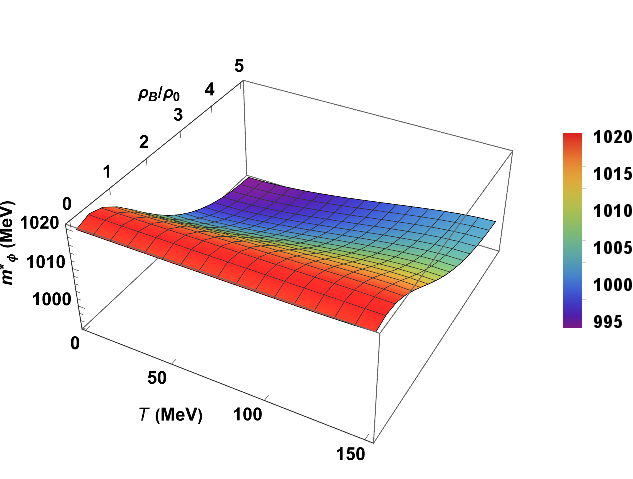}
	\end{subfigure}\hfill 
	\begin{subfigure}{.5\linewidth}
		(d)\includegraphics[width=0.9\linewidth]{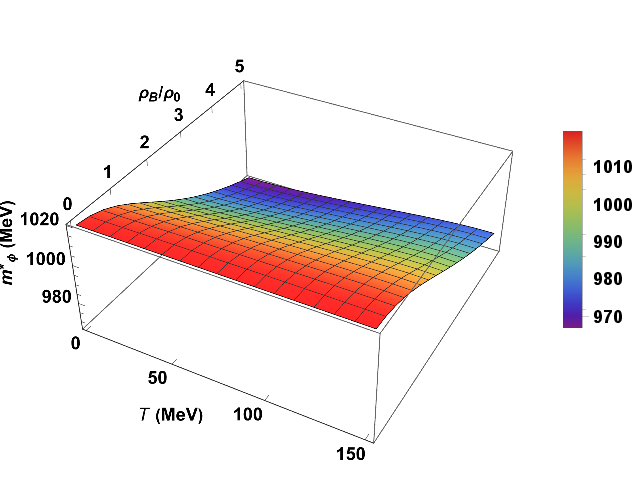}
	\end{subfigure}
	\caption{Same as Fig. \ref {3D_mphi_2CF}, at $\Lambda_c=4$ GeV.}
	\label{3D_mphi_4CF}
\end{figure}
\par
\begin{figure}
		\centering
		\includegraphics[width=0.9\linewidth]{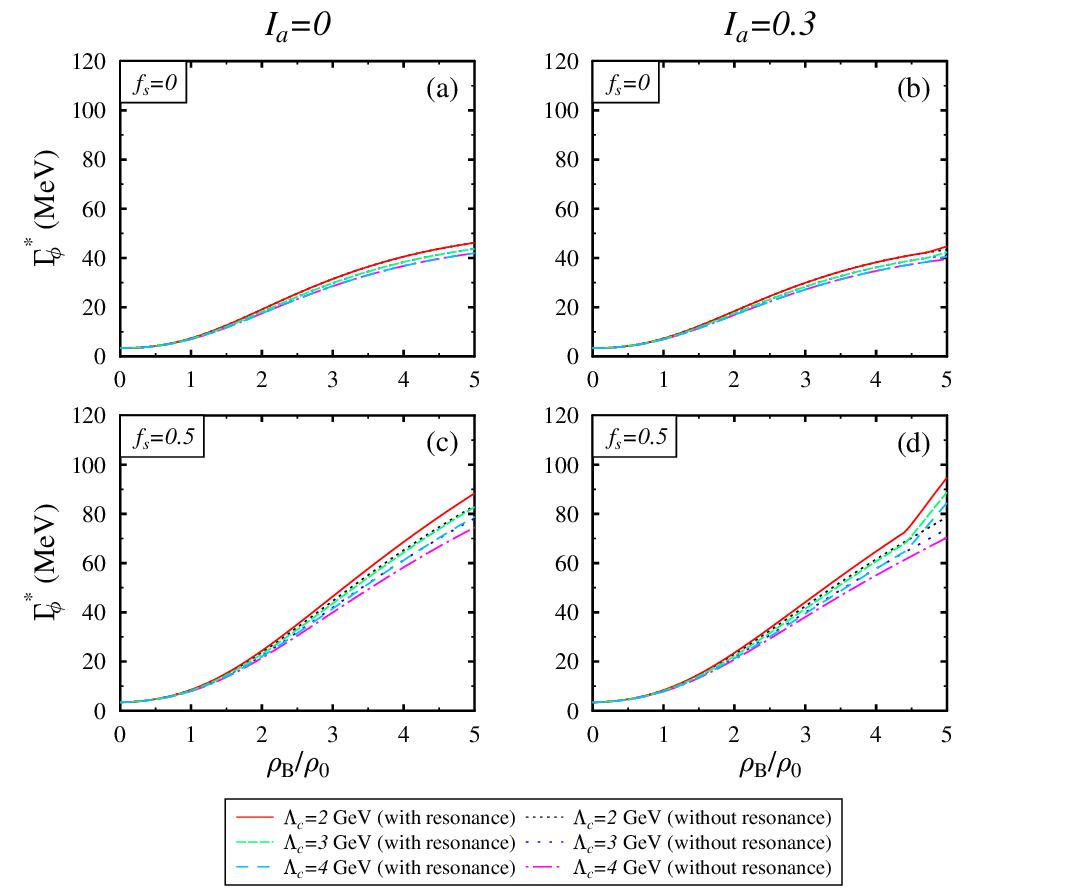}
		\caption{The in-medium decay width $\Gamma^*_{\phi}$  of $\phi$ meson  in dense resonance matter with respect to baryon density ratio $\rho_B/\rho_0$ for various values of $\Lambda_c$, $I_a$, and $f_s$ at temperature $T=0$ MeV. Each subplot also compares the cases in which decuplet baryons are not taken into account within the medium.}
		\label{phiDWt0}
	\end{figure}
	
	\begin{figure}
		\centering
		\includegraphics[width=0.9\linewidth]{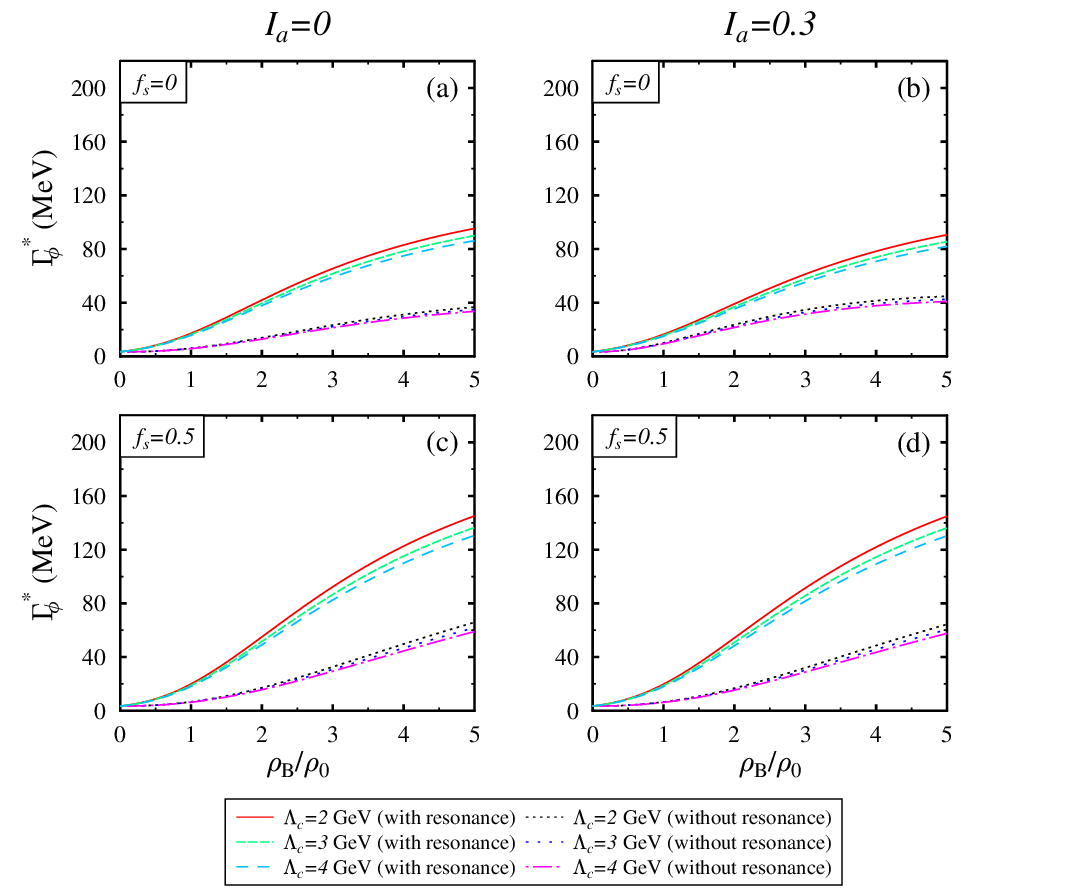}
		\caption{The in-medium decay width $\Gamma^*_{\phi}$ of the $\phi$ meson  is plotted for dense resonance matter with respect to baryon density ratio $\rho_B/\rho_0$ for various values of $\Lambda_c$, $I_a$, and $f_s$ at temperature $T=100$ MeV. These findings are compared in each subplot with the situation in which only spin-1/2 baryons are considered.}
		\label{phiDWt100}
	\end{figure}
  As we shift to a strange medium where $f_s=0.5$, it is noticed that $m^*_{\phi}$ exhibits similar behavior in both media (with and without resonance) as the function of $\rho_B/\rho_0$ ratio as well as $\Lambda_c$, as can be seen in Figs. \ref{massT0} (c) and (d). In a medium containing only hyperons, as the value of cutoff parameter $\Lambda_c$ increases, the $m^*_{\phi}$ curve shifts downward. However, when decuplets are incorporated into this medium, the same trend is observed with a more remarkable reduction in $m^*_{\phi}$ as $\Lambda_c$ increases for all baryonic densities due to dominant attractive interactions. This significant reduction in $m^*_{\phi}$ reflects the medium-modified mass behavior of kaon (antikaon) in these environments. With the increase in temperature, $T=100$ MeV, a more pronounced reduction in $m^*_{\phi}$ is observed with increasing density and $\Lambda_c$ as compared to $T=0$ MeV, as shown in Fig. \ref{massT100}.
  At $T=0$ and 100 MeV, in the non-strange ($f_s=0$) isospin asymmetric ($I_a=0.3$) resonance medium, at nuclear saturation density, $\rho_0$, average kaons (antikaons) mass shift of $22$ ($-44.5$) and $15$ ($-78.3$) MeV leads to only $-5$ MeV and $-13$ MeV mass shift of $m^*_{\phi}$,   respectively. At $f_s=0.5$, these mass shift values for $K$ ($\bar{K}$) change to $-11.5$ ($-16.7$) and $-17.7$ ($-58$) MeV inducing shift of $-6.3$ MeV and $-15.7$ MeV in $m^*_{\phi}$.
For a given baryon density $\rho_B$, mass $m^*_{\phi}$ is observed to decrease  with an increase of $\Lambda_c$, for all different environments. These findings align with results from previous studies, where in-medium $\phi$ meson masses were calculated in nuclear medium \cite{Cobos-Martinez:2017vtr, Cobos-Martinez:2017woo} and hyperonic medium \cite{kumar2020phi}. The studies in Refs. \cite{Cobos-Martinez:2017vtr} and \cite{kumar2020phi} observed that the vacuum $\phi$ meson mass (1020 MeV) falls to 994.9  and 1017.4 MeV, respectively, in nuclear medium at $\Lambda_c=3$ GeV, $\rho_B= \rho_0$, and $T=0$ MeV.
 The larger shift in $m^*_{\phi}$ observed in Ref. \cite{Cobos-Martinez:2017vtr} can be attributed to the omission of the difference between $m^*_{K}$  and $m^*_{\bar K}$ in their calculations, reporting a mass shift of roughly $-64$ MeV for both $K$ and $\bar K$ at $\rho_0$.
  At $T=0$ MeV, in Refs. \cite{Klingl:1997tm, Gubler:2014pta, Gubler:2016itj},  nearly 1\%  mass shift for $\phi$ mesons was observed in the QCD sum rules calculations.
 \par 
   In Figs. \ref{3D_mphi_2CF}, \ref{3D_mphi_3CF}, and \ref{3D_mphi_4CF} we show
   the 3D plots for  visualization of  density and temperature dependence of the in-medium $\phi$ meson masses, at $\Lambda_c =2$, 3, and 4 GeV, respectively.
 Fig. \ref{3D_mphi_2CF} (a) and (c) illustrate an isospin asymmetric nonstrange medium, with (a) includes $n,p$ and $\Delta^{++,+,0,-}$ baryons, while (c) includes only nucleons. Calculations considering only nucleons exhibit a significant decrease in $m^*_{\phi}$ as temperature drops, whereas incorporating $\Delta$ baryons, the opposite trend of $\phi$ meson mass is observed. The subplots (b) and (d) of Fig. \ref{3D_mphi_2CF} represent $f_s=0.5$ and $I_a=0.3$, with (b) involving all octet and decuplet baryons and  (d) includes octet baryons only.
  Consideration of all spin$-3/2$ resonance baryons along with the spin$-1/2$ octet in Fig. \ref{3D_mphi_2CF} (b)  produces a comparable trend to Fig. \ref{3D_mphi_2CF} (a), but with a more noticeable modification in $m^*_{\phi}$ has been observed in Fig. \ref{3D_mphi_2CF} (b) as a function of $T$ and $\rho_B/\rho_0$ ratio. As shown in Fig. \ref{3D_mphi_3CF} and \ref{3D_mphi_4CF}, an increase in the cutoff mass value results in a more significant decrease in $m^*_{\phi}$.

\begin{figure}[hbt!]
		\begin{subfigure}{.5\linewidth}
			(a)\includegraphics[width=0.85\linewidth]{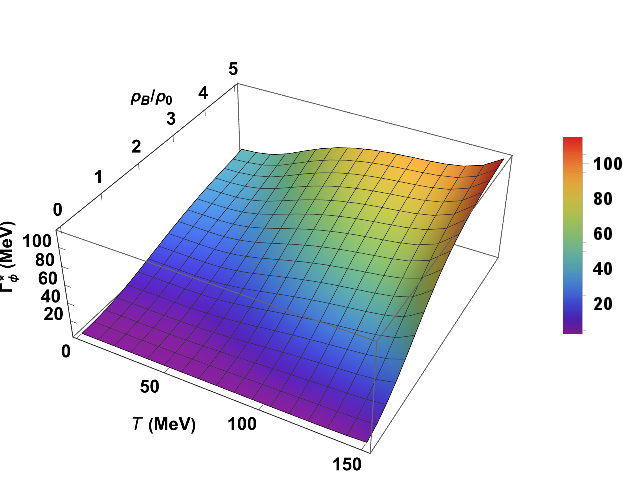}
			
		\end{subfigure}\hfill 
		\begin{subfigure}{.5\linewidth}
			(b)\includegraphics[width=0.9\linewidth]{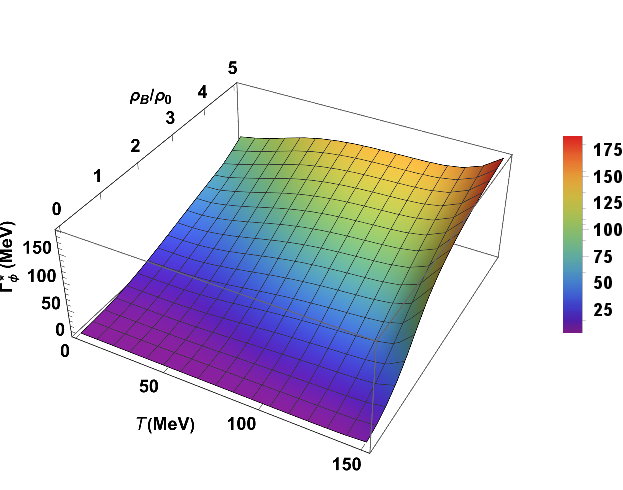}
		\end{subfigure}
		\begin{subfigure}{.5\linewidth}
			(c)\includegraphics[width=0.9\linewidth]{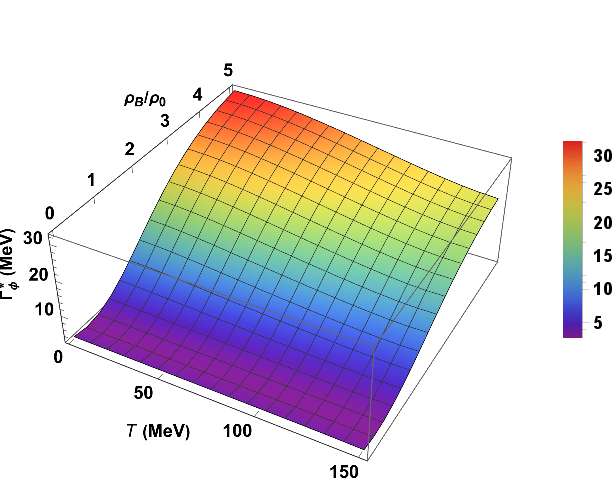}
		\end{subfigure}\hfill 
		\begin{subfigure}{.5\linewidth}
			(d)\includegraphics[width=0.9\linewidth]{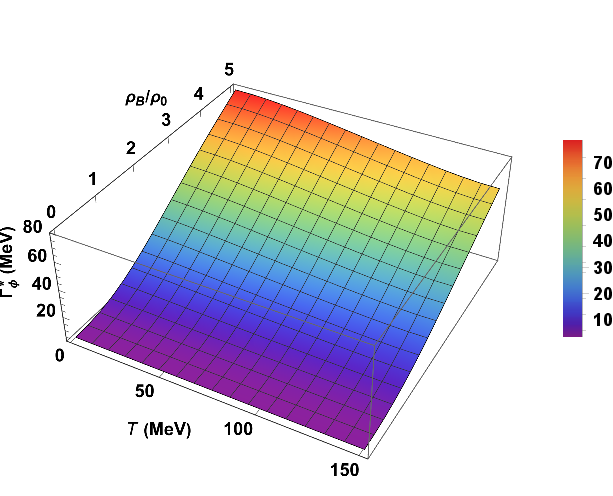}
		\end{subfigure}
		\caption{
			The 3D representation of the in-medium decay width of $\phi$ meson is plotted as a function of baryon density ratio $\rho_B/\rho_0$ and temperature $T$ for an isospin asymmetric medium ($I_a=0.3$) with the cutoff parameter $\Lambda_c=2$ GeV. The subplots (a) and (c) correspond to $f_s = 0$, with (a) incorporating both nucleons and $\Delta$ baryons while (c) involves only nucleons. The subplots (b) and (d) are for $f_s=0.5$, where (b) includes all octet and decuplet baryons and (d) involves octet baryons only.} 
		\label{3D_DW_2CF}
	\end{figure}
	\begin{figure}[hbt!]
		\begin{subfigure}{.5\linewidth}
			(a)\includegraphics[width=0.9\linewidth]{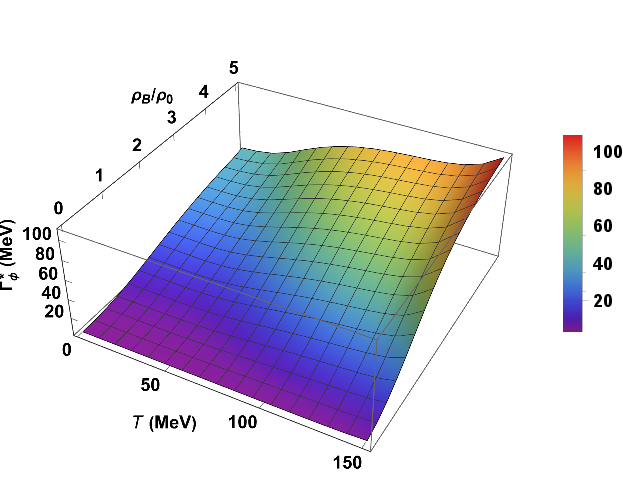}
			
		\end{subfigure}\hfill 
		\begin{subfigure}{.5\linewidth}
			(b)\includegraphics[width=0.9\linewidth]{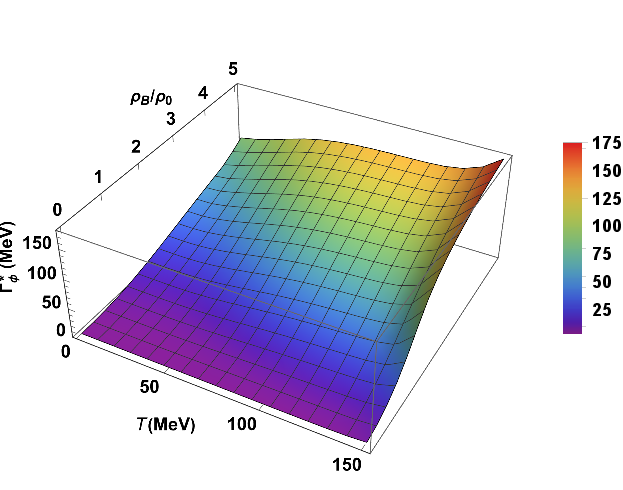}
		\end{subfigure}
		\begin{subfigure}{.5\linewidth}
			(c)\includegraphics[width=0.9\linewidth]{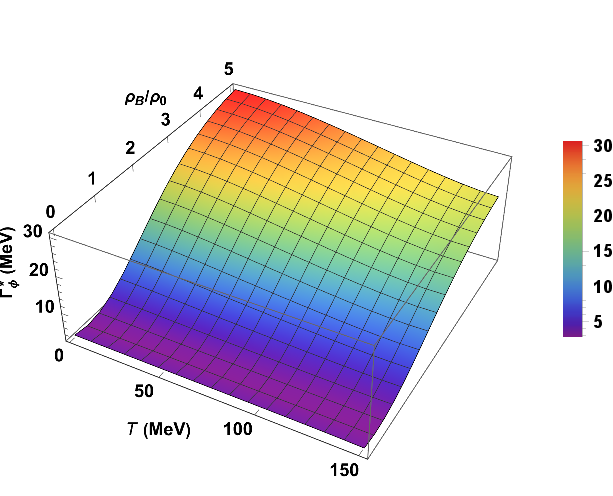}
		\end{subfigure}\hfill 
		\begin{subfigure}{.5\linewidth}
			(d)\includegraphics[width=0.9\linewidth]{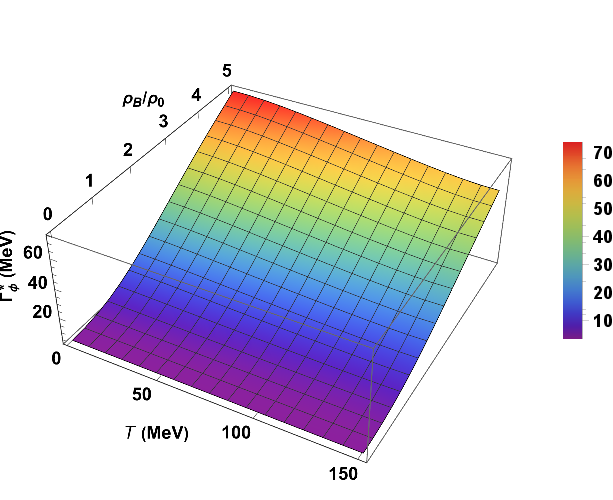}
		\end{subfigure}
		\caption{Same as Fig. \ref {3D_DW_2CF}, at $\Lambda_c=3$ GeV.}
		\label{3D_DW_3CF}
	\end{figure}
	\begin{figure}[hbt!]
		\begin{subfigure}{.5\linewidth}
			(a)\includegraphics[width=0.9\linewidth]{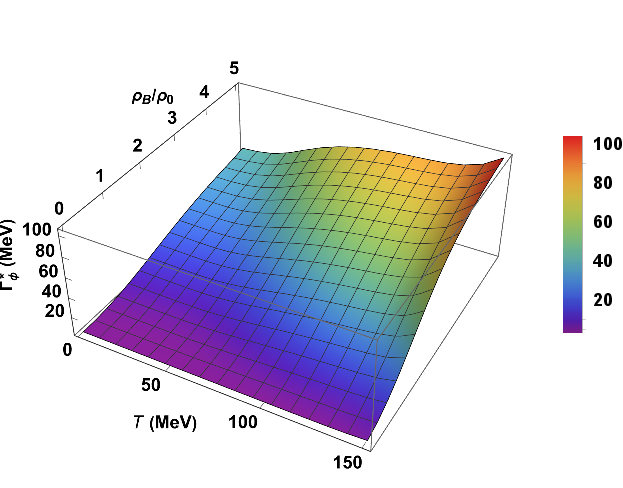}
			
		\end{subfigure}\hfill 
		\begin{subfigure}{.5\linewidth}
			(b)\includegraphics[width=0.9\linewidth]{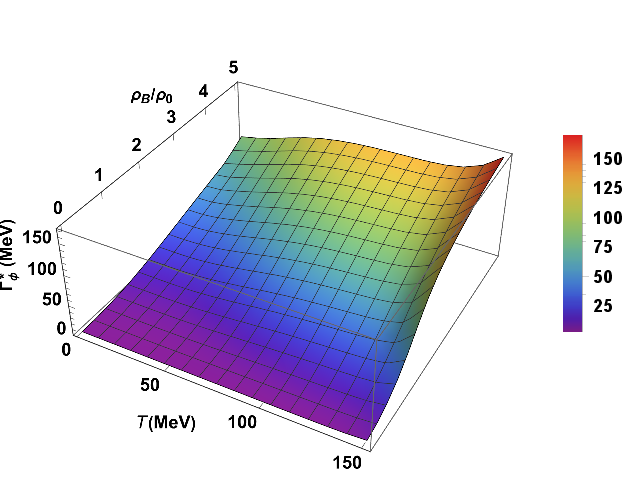}
		\end{subfigure}
		\begin{subfigure}{.5\linewidth}
			(c)\includegraphics[width=0.9\linewidth]{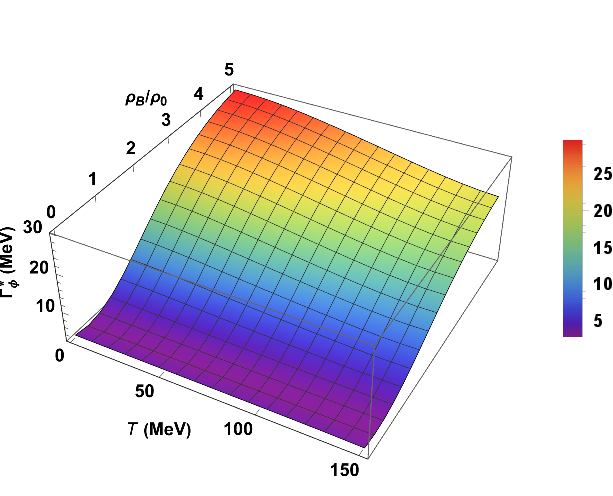}
		\end{subfigure}\hfill 
		\begin{subfigure}{.5\linewidth}
			(d)\includegraphics[width=0.9\linewidth]{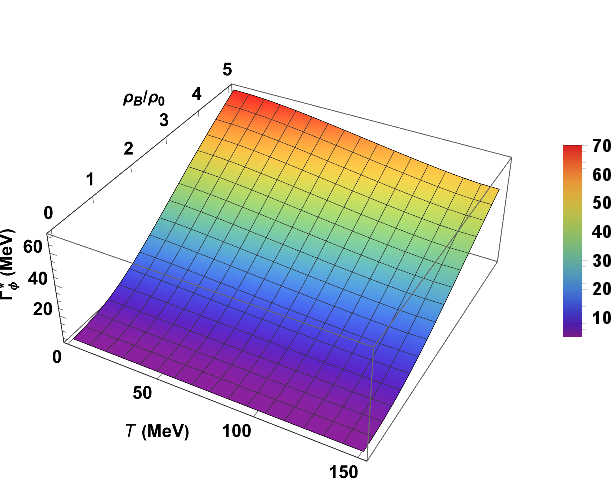}
		\end{subfigure}
		\caption{Same as Fig. \ref {3D_DW_2CF}, at $\Lambda
        _c=4$ GeV.}
		\label{3D_DW_4CF}
	\end{figure}

\subsection{The in-medium decay width of $\phi$ meson in resonance matter}
\label{results_phi_mass}
This subsection addresses how resonance production within the medium influences the probability of $\phi$ meson decay. As discussed earlier in Sec.\ref{math} C, the decay width of $\phi$ meson is derived from the imaginary part of the self-energy Im$\Pi^*_{\phi}$. Figs. \ref{phiDWt0} and \ref{phiDWt100} show the behavior of the in-medium decay width of the $\phi$ meson as a function of baryonic density in isospin-dependent resonance medium for the different values of $\Lambda_c=2, 3, 4$ GeV, at temperatures $T=0$ and 100 MeV. Table \ref{table_DW} presents the medium-modified values of decay width of $\phi$ meson $\Gamma^*_\phi$ within the resonance medium for various values of $\Lambda_c$, $\rho_B$, $I_a$, $f_s$, and $T$. We have noticed that raising the baryonic density leads to an increase in $\Gamma^*_\phi$ for all scenarios involving different values of $f_s$, $I_a$, $\Lambda_c$, and $T$. In Fig. \ref{phiDWt0} (a) and (b), the medium modification of $\Gamma^*_\phi$ as a function of $\rho_B/\rho_0$, with and without $\Delta$ baryon resonances, remains the same.
Further inclusion of strange baryons gives a more appreciable increase in $\Gamma^*_\phi$  as illustrated in Fig. \ref{phiDWt0} (c) and (d). Notably, the rate of increase in $\Gamma^*_\phi$ is more substantial in presence of resonance baryons,
 as the cutoff mass parameter increases, for all values of $\rho_B$.
In Fig. \ref{phiDWt100}, the decay width of $\phi$ meson at finite temperature exhibits a pattern similar to that observed at zero temperature. However, there is a more pronounced increase in $\Gamma^*_\phi$ with respect to baryonic density at higher temperatures. 

\begin{table}
    \centering
    
    \begin{tabular}{|c|c|c|c|c|c|c|c|c|c|c|c|c|c|}
    \hline
& & \multicolumn{4}{c|}{T=0 MeV}  & \multicolumn{4}{c|}{T=100 MeV}  & \multicolumn{4}{c|}{T=150 MeV} \\
         \cline{3-14}
       $\Gamma^*_{\phi}$ (MeV) &$f_s$ & \multicolumn{2}{c|}{$I_a=0$} & \multicolumn{2}{c|}{$I_a=0.3$} & \multicolumn{2}{c|}{$I_a=0$} & \multicolumn{2}{c|}{$I_a=0.3$} & \multicolumn{2}{c|}{$I_a=0$} & \multicolumn{2}{c|}{$I_a=0.3$} \\
        \cline{3-14}
         & &$\rho_0$&$4\rho_0$ &$\rho_0$&$4\rho_0$ &$\rho_0$&$4\rho_0$ &$\rho_0$&$4\rho_0$ &$\rho_0$&$4\rho_0$ &$\rho_0$&$4\rho_0$ \\ \hline
$\Lambda_c=2$ GeV & 0 & 7.4 & 40.5 & 7.3 & 38.2 & 17.2 & 83.0 & 16.4 & 78.3 & 30.1 & 107.1 & 30.2 & 107.5\\
\cline{2-14}
 & 0.5 & 8.5 & 68.6 & 8.4 & 64.7 & 20.0 & 122.6 & 19.8 & 121.9 & 39.0 & 163.8 & 40.3 & 167.9\\
\cline{1-14}
$\Lambda_c=3$ GeV & 0 & 7.2 & 38.3 & 7.1 & 36.2 & 16.4 & 78.3 & 15.7 & 73.9 & 28.4 & 100.9 & 28.5 & 101.2\\
\cline{2-14}
 & 0.5 & 8.2 & 64.3 & 8.1 & 60.7 & 18.9 & 115.2 & 18.8 & 114.5 & 36.7 & 154.1 & 37.9 & 158.1\\
\cline{1-14}
$\Lambda_c=4$ GeV & 0 & 7.0 & 36.8 & 6.9 & 34.7 & 15.7 & 75.0 & 15.1 & 70.8 & 27.2 & 96.6 & 27.3 & 96.9\\
\cline{2-14}
 & 0.5 & 8.0 & 61.2 & 7.9 & 57.7 & 18.1 & 110.0 & 18.0 & 109.3 & 34.9 & 147.6 & 36.1 & 151.4\\
\cline{1-14}
\end{tabular}
\caption{The in-medium values of $\Gamma^{*}_{\phi}$ at different fixed values of $\Lambda_c$, $\rho_B$, $I_a$, $f_s$, and $T$. }
\label{table_DW}
\end{table}
 We portray a 3D representation of the decay width of $\phi$ mesons as a function of baryon density ratio $\rho_B/\rho_0$ and temperature $T$, at fixed values of $\Lambda_c =2$, 3, and 4 GeV in Figs. \ref{3D_DW_2CF}, \ref{3D_DW_3CF}, and \ref{3D_DW_4CF}, respectively. 
Within a pure nuclear medium at low temperature, there is a rise in $\Gamma^*_\phi$ with increasing density, while at a higher temperature, the rate of increase becomes slower at $\Lambda_c=2$ GeV as exhibited in Fig. \ref{3D_DW_2CF} (c). When we involve $\Delta$ baryons within the nuclear medium, the rate of increase becomes faster at higher temperatures as shown in Fig. \ref{3D_DW_2CF} (a). Furthermore, in Fig. \ref{3D_DW_2CF} (d), we observed a comparable trend of $\Gamma^*_\phi$ as in Fig. \ref{3D_DW_2CF} (c) because these both Figures considered spin 1/2 baryons only. Fig. \ref{3D_DW_2CF} (b), involves all spin 1/2 and 3/2 baryons, the rate of increase in $\Gamma^*_\phi$ becomes notably more rapid as temperature of the medium increases.  
Figs. \ref{3D_DW_3CF} and \ref{3D_DW_4CF} show that as $\Lambda_c$ increases at given $\rho_B/\rho_0$, the rate of increase in $\Gamma^*_\phi$ exhibits a decreasing trend in both the resonance and non-resonance cases, as can be seen in Table \ref{table_DW_ref}, where the comparison of $\Gamma^*_\phi$ of the nuclear and resonance matters is listed at $T=0$ and 100 MeV for $I_a=0.3$.
\begin{table}[h]
	\centering
	
	\begin{tabular}{|c|c|c|c|c|}
		\hline
		\multirow{2}{*}{$\Gamma^*_\phi$ (MeV)} & \multicolumn{2}{c|}{Nuclear Matter} & \multicolumn{2}{c|}{Resonance Matter} \\
		\cline{2-5}
		& $T=0$ MeV       & $T=100$ MeV  & $T=0$ MeV & $T=100$ MeV \\
		\hline
		$\Lambda_c = 2$ GeV     & 7.30    & 10    & 8.4     & 19.8\\
		\hline
		$\Lambda_c = 3$ GeV  & 7.09  & 9.7   & 8.1   & 18.8 \\
		\hline
		$\Lambda_c = 4$ GeV   & 6.93  & 9.4 & 7.9 & 18.0\\
		\hline
\end{tabular}
\caption {The values of $\Gamma^*_\phi$ (MeV) for different $\Lambda_c$ in isospin asymmetric nuclear and resonance matter at $T=0$ MeV and $T=100$ MeV at nuclear saturation density, $\rho_B=\rho_0$.}
\label{table_DW_ref}
\end{table} 
The findings from the present study align with the outcomes reported in Refs. \cite{Cobos-Martinez:2017vtr}, indicating a reduction in decay width when $\Lambda_c$ is altered from 2 to 4 GeV.
Ref. \cite{Cabrera:2002hc} examines the self-energy of $\phi$ for a symmetric nuclear environment. This analysis involves incorporating the self-energies of $S$ and $P$-waves. The self-consistent coupled channel calculation, based on effective chiral Lagrangians was employed to determine the $S$-wave $\bar K$ self-energy.
The $P$-wave component is influenced by hyperon-hole excitation couplings. The results indicate $\phi$ meson mass shift of approximately 8 MeV and a width of about 30 MeV in the nuclear medium at nuclear saturation density.


\section{Summary}
 To conclude, we examine the in-medium properties of the $\phi$ meson in isospin asymmetric resonance medium through the medium modifications of $K$ and $\bar K$ meson masses. Employing a mean-field approximation within the chiral SU(3) framework, we compute the medium-modified properties of baryons by exchange of scalar fields, $\sigma, \zeta$ and $\delta$ and vector fields, $\omega, \rho$, and $\phi$.
 Vector and scalar densities of baryons and in-medium scalar meson fields 
 $\sigma, \zeta$ and $\delta$ serve as input to analyze the behavior of effective masses of $K$ and $\bar K$ mesons in a dense resonance environment. 
 The medium-modified masses of kaons and antikaons are further utilized to determine the $\phi$ meson masses and decay width in isospin asymmetric resonance matter. Considering the presence of resonance baryons,  we observe attractive interactions (negative mass shift) for kaon and antikaon masses at higher baryon density and temperature of the medium.
   Due to this the decay width of the $\phi$ meson is observed to increase in presence of resonance baryons. 
At temperature $T=100$ MeV, the effective masses  of kaons and antikaons are observed to decrease by 3.6\% and 11.8\%, respectively.
This  results in a 1.56\% reduction in the mass of the $\phi$ meson within an isospin asymmetric resonance medium at nuclear saturation density. 

\par
As previously mentioned, the in-medium properties of vector mesons are utilized as input in various transport models to simulate and predict the results of different experiments \cite{Song:2022jcj, Pal:2002aw, Weil:2011fa, Barz:2009yz}.
The future experiment E16 at J-PARC examines dielectron spectra to investigate how vector mesons spectral properties change within the medium, which indicates a partial restoration of broken symmetry. Furthermore, this experiment will measure the $\phi$ meson spectrum at both zero and finite momentum \cite{Aoki:2023qgl}. In practical applications, the decay width serves as a valuable tool for evaluating $\phi$ meson creation in $p-N$ collisions. The production, absorption, and momentum spectrum of $\phi$ meson within the hadronic environment are expected to be addressed through the findings of experiments like CBM and PANDA at FAIR  \cite{senger2012compressed,rapp2010charmonium, Destefanis:2013xpa}.
\label{summary}
\section*{Acknowledgment}
The authors sincerely acknowledge the support for this work from the Ministry of Science and Human Resources (MHRD), Government of India, through an Institute fellowship under the National Institute of Technology Jalandhar.
Arvind Kumar sincerely acknowledge 
Anusandhan National Research Foundation (ANRF),
Government of India for funding of the research project under the
Science and Engineering
Research Board-Core Research Grant (SERB-CRG) scheme (File No. CRG/2023/000557).
\bibliography{ref_phi}

\end{document}